\documentclass[a4paper,12pt]{article}
\usepackage[latin1]{inputenc}
\usepackage[english]{babel}
\usepackage{amssymb}
\usepackage{amsfonts}
\usepackage{eufrak}
\usepackage{subfig}

\usepackage[pdftex]{graphicx}
\usepackage[margin=2.5cm]{geometry}

\usepackage{hyperref}
\title{Some properties of the Higgs sector of the Next-to Minimal SuperSymmetric Model}
\author{Jacob Winding}

\begin{document}

\begin{flushright}
LU TP 11-22 \\
June 2011
\end{flushright}
{\centering{ \large \bf Bachelor thesis\\ }}
\vspace{ 1 cm}
{\centering \huge \bf Some properties of the Higgs sector of the Next-to Minimal SuperSymmetric Model \\}
\vspace{1 cm}
{\centering \large Jacob Winding\\}
\vspace{1 cm}
{\centering  Theoretical High Energy Physics\\ Departement of Astronomy and Theoretical Physics \\Lund University\\S\"olvegatan 14A \\ SE-223 62 Lund\\} 
\vspace{1.5 cm}

\begin{center}
{\large 
Supervised by: Johan Rathsman}
\end{center}
\vspace{3 cm}
\begin{abstract}
The problems of the standard model are briefly reviewed and the motivations for introducing supersymmetry are discussed. Two realistic supersymmetric models; the Minimal SuperSymmetric Model, MSSM, and its proposed extension NMSSM are introduced briefly and then some details of the Higgs sector of NMSSM are stated and some constraints on parameters are described. I then explore the Higgs boson masses and couplings for some interesting scenarios, including a few different ways of taking the limit where NMSSM reduces to MSSM.
\end{abstract}

\newpage

\section{Introduction}

The standard model, despite being very successful and in overall very good agreement with experiments, has many obvious shortcomings. One of the leading contenders claiming to solve some of these problems is supersymmetry (SUSY). As the perhaps most important example, we have the problem of the small Higgs mass in the standard model, called the \emph{hierarchy problem}. This problem is automatically solved in supersymmetric theories. From a theoretical viewpoint, supersymmetry is attractive not only because it solves some problems, but also since it can be said to be the most general possible extension of our normal spacetime symmetries\cite{ColemanMandula, Haag:1974qh}.

In this study, which is part of a Bachelor thesis at Lund University, some problems of the standard model and how they motivate the study of supersymmetry are reviewed. Then I briefly introduce the minimal supersymmetric model, MSSM, and one of its proposed extensions with a singlet Higgs field, the next-to MSSM or NMSSM. Then I examine the Higgs sector of NMSSM in some more detail, and look at some of its qualitative features, using numerical tree level calculations to look at the mass spectrum and the reduced couplings. Some different limits where NMSSM reduces to MSSM are also studied.

\subsection{Supersymmetry and some problems of the standard model}
The standard model, while highly successful and in good agreement with the absolute majority of performed experiments (there are however some persistent measured deviations, like the anomalous magnetic moment for the muon\cite{ParticleDataGroup}), still has some serious problems and is theoretically very unsatisfactory. For one thing, it has a large number (at least 19) of arbitrary parameters, including the particle masses, the three gauge couplings, the weak mixing angle and the CP-violating Kobayashi-Maskawa phase. This is a problem which really isn't improved much by supersymmetry (in fact the breaking of supersymmetry seems to introduce a lot of new parameters).

Another obvious deficit of the standard model is that it doesn't describe the fourth known force, gravity. What one would ultimately want is a theory that describes all the known forces in a unified way. Since supersymmetry is an extension of the ordinary spacetime symmetries rather than a new internal symmetry, it is natural how curved spacetime and thus gravity follows from making supersymmetry local. The resulting theory is called \emph{supergravity}, see \cite{supergravity} for an introduction. This doesn't really give us a realistic and consistent quantum theory of gravity but it at least might hint at how this may be achieved.

The standard model also has to be modified in order to be consistent with the standard theories of cosmology. Especially, it cannot explain the observed cold dark matter, and when one calculates the vacuum energy in the standard model it gives contributions that are far too large to match the observed small, but nonzero, value of the cosmological constant. Supersymmetry offers a good candidate for dark matter, since it turns out that the lightest supersymmetric partner (which probably is a mixture of the superpartners of the photon, Z-boson and neutral Higgs bosons, called the neutralino) has to be stable, if we assume R-parity (which in turn is strongly implied from limits on the proton lifetime).  

Then there is the hierarchy problem concerning the smallness of gravity compared to electromagnetism. Since gravity couples to mass, this problem is equivalent to asking why the particle masses are so much smaller than the Planck mass scale ($M_P\sim 10^{19}$ GeV), i.e.\ why the ratio of the electroweak scale ($\sim M_W$) and the Planck scale is so tiny, $M_W/M_P \sim 10^{-16}$. From reasons stated above, we know that the standard model is an effective theory, at most valid up to the Planck scale. This means that when renormalizing the theory we must have a finite energy cutoff, that can at largest be the Planck mass. This finite cutoff is really what causes the problem. The masses of the particles come from the Higgs mechanism, so the question why the masses of the particles are small is in turn equivalent to asking why the Higgs has such a small mass. If one calculates loop corrections to the Higgs propagator, one finds that the corrections look like 
\[ \delta m^2_H \eqsim \mathcal{O}(\alpha)\Lambda^2, \]
so the correction to the mass squared are proportional to the square of the UV-cutoff $\Lambda$, which by our previous reasoning is some large, finite energy, maybe of order $M_P$ or at least of some large unification scale (called the GUT-scale, for Grand Unified Theory). So in order to give the Higgs its required mass, which of course is much smaller than either of these scales, these loop corrections has to be very precisely cancelled. This we can do by giving the tree level diagram exactly the required value. Technically, this is not a problem since there is no constraint on the value of the bare mass, but it introduces a very heavy fine-tuning into the theory, where the bare mass has to have exactly the correct value, and this is theoretically very unsatisfying. 
 
A thing to note is that implicit in this line of reasoning is that you assume that there is no need for new physics below some very large energy, i.e.\ the cutoff scale is large. This is called the ``big desert'' assumption, and while many, holds this to be true, not all physicists agree. An argument for this assumption is that in the standard model, the running of the couplings is such that at a high energy scale, the grand unified scale (GUT-scale), of about $\sim 10^{16}$ GeV, all the known coupling become roughly the same, which implies that at least at this scale, our current physics should drastically change. Supersymmetry actually improves this a bit, making the couplings meet more closely than in the standard model. If one accepts the big-desert assumption, then the hierarchy problem is real, and some mechanism is needed to keep the Higgs mass small. Supersymmetry is the most popular proposal to solve this problem, but other theories exists, such as technicolor \cite{Weinberg1976,SusskindTechnicolor} (in which the Higgs is a composite particle) and extra dimensions (for example the ADD model\cite{ADD}). 

A good reason to study supersymmetry is therefore that it solves this hierarchy problem in a natural way. It is solved since for every fermion that couples to the Higgs, SUSY adds a scalar with the same quantum numbers. When calculating the loop corrections, the fermion loops and the scalar loops will be of the opposite sign and (if supersymmetry wasn't broken) be of the same size and thus cancel. Even when supersymmetry is broken, these cancellations removes the dependence on $\Lambda^2$ and reduces it to a logarithmic divergence\cite{Kaul}, if supersymmetry is softly broken.

A final problem of the standard model worth mentioning briefly is the strong CP problem. This comes from the observational fact that the strong interaction as described by QCD doesn't seem to violate CP symmetry, in contrast with the weak force. This is a problem since there are natural terms in the QCD Lagrangian that violates CP conservation. To conform to the experimental data, a large amount of fine-tuning is again required. The most well-known proposition for solving this problem introduces new scalar particles, called axions\cite{PQ}, which can make an appearance in the NMSSM, as discussed later. 

Supersymmetry is a vast field, and the theory behind it is an exciting subject. For more complete expositions of SUSY and its consequences, there are numerous books and articles to consult, for example \cite{Ellis:2002mx:AlpHikers,Lykken:1996xt,FigueroaO'Farrill:2001tr,Gates:1983nr:superspace1001}.

\section{Realistic supersymmetric models}
In this section I will introduce the simplest way to extend the standard model into a supersymmetric theory, the minimal supersymmetric standard model (MSSM). Then an extension to this theory is presented, the next-to MSSM (NMSSM), mostly in order to solve a problem concerning the value of a dimensionfull parameter in the MSSM. Since nature obviously isn't supersymmetric at low energies, we also need to study how supersymmetry is broken. This is a large subject which I won't cover in any detail, only introduce the concept of how we can introduce so called softly breaking terms into the Lagrangian of our models, as discussed in the next section.

\subsection{Softly breaking terms}\label{softlybreakingterms}
The most popular ideas about how supersymmetry breaking works, is that it is spontaneously broken, in a manner similar to how the gauge symmetries are broken, see \cite{Shadmi:2006jv} for a discussion. In fact, supersymmetry is broken as soon as the vacuum gets a nonzero energy. This fact means that the breaking of supersymmetry and the breaking of gauge symmetries are closely connected subjects. There are different additional terms you can introduce into your Lagrangian, such that you can give the fields in these terms a non-vanishing VEV, and thus give the vacuum a nonzero energy, breaking supersymmetry. For more details, see \cite{Shadmi:2006jv}.

When constructing realistic models, we don't really need to care about the details of exactly how this happens. Instead, we can introduce terms into our Lagrangian that explicitly breaks supersymmetry, but at the same time preserves renormalizability and are such that at high energy, above the supersymmetry breaking scale, they become irrelevant. Such terms are called \emph{softly breaking terms}, and are essentially things like scalar mass terms, gaugino masses or cubic scalar terms with dimensionfull couplings. For our purposes, analysing the Higgs sector at leading order, we only care about the scalar mass terms. So when we have the supersymmetric Lagrangian, we can then add all such allowed terms, and view them as an effective description of how supersymmetry is broken.

\subsection{The Minimal SuperSymmetric Model}
Just as it sounds, the Minimal SuperSymmetric Model (MSSM) is the model you get when you try to minimally extend the standard model to incorporate supersymmetry. Since none of the particles in the standard model have the same quantum numbers (excluding mass), one cannot let any of the known particles be each others superpartners. So instead we let every particle be a part of a corresponding superfield, and then put the superfields in the same $SU(2)_L$ doublets as in the SM. The same of course applies to the gauge fields, which now become part of gauge superfields.  

In a supersymmetric theory, only a Higgs with hypercharge $Y=1/2$ can have the necessary Yukawa coupling to give masses to the up-type quarks with charge $+2/3$, and only a Higgs with hypercharge $-1/2$ can have the necessary couplings to give mass to the down-type quarks. This is because the superpotential is holomorphic, so the Higgs doublet giving mass to the up-type quarks cannot also give mass to the down-type quarks since we are not allowed to use the complex conjugate. Thus we at least need two different Higgs $SU(2)_L$-doublets in order to give mass to all the massive particles. We will call the $Y=1/2$ doublet $H_u$, and the $Y=-1/2$ doublet $H_d$. Then the upper component of the $Y=1/2$ will have isospin $T_3=+1/2$ and therefore have electric charge $+e$. The lower component will be electrically neutral, and in the same way the upper component of the $Y=-1/2$ doublet will be neutral while the lower will have a negative electric charge. So the superdoublets look like
\begin{equation}
\hat H_u = \left(\begin{array}{c}\hat H_u^+\\\hat H_u^0\end{array}\right), \hspace{5 mm} \hat H_d = \left(\begin{array}{c}\hat H_d^0\\\hat H_d^-\end{array}\right)
\end{equation}
and the Higgs fields which gives masses to the fermions will be the corresponding scalar fields. We can then note that the Higgs superdoublet $\hat H_d$ has the same quantum numbers as left handed leptons (and sleptons). Therefore we can use it to give mass to the leptons as well as the down-type quarks, so we don't need another Higgs doublet for this purpose. There is another way to motivate the need of two different Higgs doublets which is based on anomaly cancellations, but this isn't logically needed.

The minimal superpotential involving these superfields which in a reasonable way extends the standard model is
\begin{equation}
 W_{MSSM} =\sum_{i,j} y^{ij}_u \hat u_i \hat H_u\cdot \hat Q_j  - y^{ij}_d \hat d_i \hat H_d\cdot \hat Q_j  -y_e^{ij} \hat e_i \hat H_d\cdot \hat L_j + \mu \hat H_u\cdot \hat H_d
\end{equation}
where the ``hatted'' letters denote the superfield doublets or singlets corresponding to the normal $SU(2)_L$ doublets/singlets in the standard model, and $i,j$ are generation indices, $i=1,2,3$. That is, for the first generation, $\hat Q_1 = ( \hat u, \hat d)^T$, $\hat L_1 = (\hat e_L, \hat \nu_e)^T$, $\hat e_1 = \hat e_R, \hat u_1 = \hat u$ and so on. The Higgs doublets are as described above, and the $\mathbf y^{ij}$ are the Yukawa couplings among generations. The products of $SU(2)_L$ doublets are given by 
\[ A\cdot B = \epsilon_{ab} A^a B^b \]
where $\epsilon_{ab}$ is the fully antisymmetric symbol in two dimensions with $\epsilon_{12}=1$ and $a,b$ are $SU(2)_L$ indices. In this superpotential the $\mu$ parameter has dimension mass and is contributes to the masses of Higgs fields, and it is this simple fact which motivates the introduction of the next-to minimal supersymmetric standard model (NMSSM). 

From this superpotential and the ordinary gauge couplings of the standard model, we can calculate the scalar potential, by calculating the F and D contributions in the usual way. Doing this, and looking only at the Higgs sector of the potential, we find
\begin{equation}
V_F = \mu^2 (|H_u|^2 + |H_d|^2)
\end{equation}
and 
\begin{equation}
V_D = \frac 1 8 g^2 \left (|H_u|^2 - |H_d|^2 \right )^2 + \frac 1 2 g_2^2 |H_u^\dag \cdot H_d |^2.
\end{equation}
As described above, we can then add the soft supersymmetry breaking terms;
\begin{equation}
V_{\rm{soft}} = m^2_{H_u} |H_u|^2 + m^2_{H_d} |H_d|^2 + (m^2_3 H_u \cdot H_d + \rm{h.c.}) 
\end{equation}
where the dimensionfull parameters $m^2_{H_u}, m^2_{H_d}$ and $m^2_3$ clearly have to be of the order of the weak or supersymmetric breaking scale. The total scalar potential is then the sum of these three terms. By letting at least one of  $m^2_{H_u}$ and $m^2_{H_d}$ be negative, $H_u$ and $H_d$ acquires non-zero VEVs, breaking the symmetry.

From requiring vacuum stability we get some relations between $m^2_{H_u}, m^2_{H_d}, m^2_3$, the VEVs and $\mu$; as described in more detail in the next section for the NMSSM. Using these, one can calculate and then diagonalize the mass matrices that describe the physical mass eigenstates in terms of the parameters of the model.

In the MSSM, it turns out that we get one physical charged Higgs state, $H^\pm$, with a mass
\begin{eqnarray}
m^2_{H^\pm} = \left ( \frac{2m_3^2}{v_u v_d} + \frac 1 4 g^2_2 \right ) \frac{v^2}{2}
\end{eqnarray}
where $\langle H_u^0 \rangle = v_u/\sqrt{2}$ and $\langle H_d \rangle = v_d/ \sqrt{2}$, i.e.\ the VEVs, and $v^2 = v_u^2 + v_d^2$, which corresponds to the VEV of the Higgs in the standard model. We also get one neutral, pseudoscalar (CP-odd) Higgs, called $A$, with a mass 
\begin{equation}
m_A^2 = \frac{2m_3^2}{\sin 2 \beta}, 
\end{equation}
where the useful angle $\beta$ is defined from $\tan \beta = \frac{v_u}{v_d}$. Finally, we also get two neutral scalar (CP-even) Higgses, $H_1$ and $H_2$ (where $H_1$ is lighter than $H_2$), that have the masses 
\begin{eqnarray}
m^2_{H_1, H_2} = \frac 1 2 \left [ m^2_A + M_Z^2 \mp \sqrt{(m_A^2 +M_Z^2)^2 - 4M_Z^2 m_A^2 \cos^2 2\beta} \right ].
\end{eqnarray}
If we remember the relation $M_W^2 = \frac 1 4 (v^2/2) g_2^2$ \footnote{The extra factor $1/2$ coming from the factors of $\sqrt{2}$ in my definition of $v_u$ and $v_d$.}, and express $\sin 2\beta$ in terms of $v_u$ and $v_d$, we see that
\begin{equation}
m^2_{H^\pm} = m_A^2 + M_W^2.
\end{equation}
We can also conclude that 
\begin{equation}
m^2_{H_1} + m^2_{H_2} = m^2_A + M_Z^2,
\end{equation}
and the more striking inequality
\begin{equation}
m_{H_1} < \min (m_A, M_Z), 
\end{equation}
meaning that no matter how we choose our parameters, $m_{H_1} < M_Z$. This is a tree level prediction, and loop corrections can lift the mass of $H_1$ above the so far established limits, but this is still an important prediction of MSSM. 

As we are excluding larger and larger values of $m_{H_1}$, this is really a problem, called the little hierarchy problem. Just as the general hierarchy problem, this concerns the separation of mass scales, because in order to generate the large loop corrections needed to increase $m_{H_1}$, the other sparticle masses needs to become very large, again creating a new unexplained mass scale in the theory. The maximal bound one can get without adding new dynamics to the theory is something like $m_{H_1} \lesssim 135$ GeV. Thus results that exclude a Higgs mass lighter than that will exclude the whole of MSSM. 

As mentioned above, the $\mu$ and $m_3^2$ parameters are dimensionfull. The $m^2_3$ parameter isn't a problem since it enters as one of the softly breaking terms, but $\mu$ enters through the ordinary Lagrangian, so the only natural values for it before SUSY breaking occurs, is either 0 or something similar to the Planck mass $M_P$. However, to be phenomenologically viable we must have a $\mu$ that is of similar size to the electroweak scale. Otherwise there would have to be miraculous cancellations between $\mu^2$ and the soft supersymmetry breaking terms. It is in order to solve this problem we motivate the study of the NMSSM, where by adding a singlet Higgs field and coupling it to the Higgs doublets, the $\mu$ parameter is generated through supersymmetry breaking. This breaking gives the singlet and the two Higgs doublets VEVs, and thus the singlet-doublet-doublet coupling gives us an effective $\mu$. This explains why $\mu$ should be roughly the same scale as the electroweak breaking scale.

For a more complete discussion of MSSM, and some discussion about how it may be discovered, see for example \cite{Martin:1997ns}.

\subsection{The Next-to Minimal SuperSymmetric Model} \label{NMSSM-section}
As stated above, the NMSSM \cite{Ellwanger:2009dp,Ellwanger:1993xa} is a proposed extension of the MSSM, which in a natural way solves the $\mu$-problem. In order to get rid of the dimensionfull $\mu$ parameter, we add a new Higgs $SU(2)_L$ singlet $\hat S$ to the theory. Of course, in principle nothing forbids a $\mu$ term just because we add a new singlet, but we take it to have the ``natural'' value 0. The new superpotential looks like
\begin{equation}
W_{NMSSM} = W_{MSSM} + \lambda \hat S \hat H_u \hat H_d + \frac 1 3 \kappa \hat S^3, 
\label{NMSSM-Lagrangian}
\end{equation}
where $\lambda,\kappa$ are new, dimensionless parameters of the model. From this superpotential and the usual gauge couplings, the F and D part of the potential can be computed. The result looks very much like in MSSM, but with some extra terms;
\begin{eqnarray}
V_F &=& |\lambda S |^2 ( |H_u|^2 + |H_d|^2) + |\lambda H_u \cdot H_d + \kappa S^2|^2 \\
V_D &=& \frac{1}{8}g^2( |H_u|^2 -|H_d|^2 )^2 + \frac 1 2 g_2^2 |H_u^\dag \cdot H_d |^2
\end{eqnarray}
Since supersymmetry has to be broken, and we don't know nor care about the details of the breaking mechanism, we also have to add to the potential all possible terms which breaks supersymmetry in the acceptable, soft way explained in section \ref{softlybreakingterms}. This soft potential looks like
\begin{equation}
V_{\rm soft} = m_{H_u}^2 | H_u |^2 + m_{H_d}^2|H_d|^2 + m_S^2|S|^2 + \left ( \lambda A_\lambda S H_u \cdot H_d + \frac 1 3 \kappa A_\kappa S^3 + \mbox{h.c.} \right ).
\end{equation}
Nothing is preventing us from adding the $m_3^2 H_u \cdot H_d$ term present in the MSSM case, but this would add an additional parameter with mass dimension, in conflict with the philosophy behind NMSSM, so we consider only the case $m_3^2 = 0$.
The entire Higgs potential is then given by the sum of these,
\begin{equation}
V_{\rm Higgs} = V_F + V_D + V_{\rm soft}.
\end{equation}
Then, as in the breaking of electroweak symmetry, we assume that $m_S^2 < 0$ so that $S=0$ is an unstable state. If we define $ \langle S \rangle =v_s/\sqrt{2} $ we see that when we expand the singlet field around its VEV we get a term $\mu = \lambda v_s/\sqrt{2}$ with mass dimension\footnote{Note that this definition of $\mu$ is a convention, which differs by a factor $\frac{1}{\sqrt{2}}$ from the most common one.}. Since this term comes from supersymmetry breaking, it's natural for it to have a value of magnitude $|\mu| \lesssim M_{\rm{SUSY}}$, where $M_{\rm{SUSY}}$ is the scale where supersymmetry is broken. This is the way in which NMSSM solves the $\mu$-problem of MSSM.

Further, we also assume that at least one of the other Higgs mass parameters $m_{H_u}^2$ and $m_{H_d}^2$ are negative, so that $H_u, H_d$ also get nonzero VEVs, as required to break the electroweak symmetry. We then have the gauge freedom to choose $\langle H_u^+ \rangle = \langle H_d^-\rangle =0$, so that the vacuum is uncharged. In this treatment, I will discuss the vacuum obtained by further assuming all the remaining VEVs to be real, and described by
\begin{equation}
\langle H_u \rangle =\frac{1}{\sqrt{2}} \left(\begin{array}{c}0\\v_u\end{array}\right), \mbox{    } H_d =\frac{1}{\sqrt{2}}  \left(\begin{array}{c}v_d\\0\end{array}\right),\hspace{4 mm} \langle S \rangle = \frac{1}{\sqrt{2}}  v_s.
\end{equation}
We then require this vacuum to be a stable local minimum of the potential, giving us three different relations of the type 
\[ \left .\frac{ \partial V }{ \partial S} \right |_{\rm{vacuum}} = 0 \]
relating the squared masses of the Higgs fields to the VEVs and the other parameters in the theory. The derivatives w.r.t. fields with zero VEVs are trivially zero. If solved for the masses, these three relations are
\begin{eqnarray}
m_u^2 &\equiv& m^2_{H_u} + |\mu|^2 = \frac{1}{8} g^2(v_d^2 - v_u^2) + \lambda \frac{v_s v_d}{2v_u}\left( \sqrt{2} A_\lambda + v_s \kappa \right) - \frac{1}{2}\lambda^2 (v_d^2 + v_s^2)\\
m_d^2 &\equiv& m^2_{H_d} + |\mu|^2 = \frac{1}{8} g^2(v_u^2 - v_d^2) + \lambda \frac{v_s v_u}{2v_d}\left( \sqrt{2} A_\lambda + v_s \kappa \right) - \frac{1}{2}\lambda^2 (v_u^2 + v_s^2)\\
m_S^2 &=& - \frac 1 2 v^2 \lambda^2 - v_s^2 \kappa^2 + \frac{1}{\sqrt{2}} A_\lambda \lambda \frac{v_u v_d}{v_s} + v_u v_d \kappa \lambda - \frac{1}{\sqrt{2}}A_\kappa v_s \kappa. \label{m_sExpression}
\end{eqnarray}
These new masses ($m_u, m_d$) are defined since when we give the singlet a VEV, effectively there will be an additional massterm of $|\mu|^2$ for the doublet fields, so we calculate the conditions for these effective masses.

We now see that a full specification of the Higgs sector in the NMSSM requires six parameters: $\lambda,\, \kappa,\, A_\lambda,\, A_\kappa ,\, \tan \beta $ and $v_s$. Conventions can be chosen such that $\lambda, \tan \beta$ and $v_s$ (and thus $\mu$) are positive, and this is what I will do. For my purposes I also keep $\kappa > 0$, since switching this sign doesn't change any of my results. In my numerical studies I will replace $A_\lambda$ by the physical mass of the charged Higgs, $m_{H^\pm}$, which of course must be positive, and we will see that the requirement of positive masses squared restrics $A_\kappa$ to the negative range.

 The ``MSSM limit'' can be approached smoothly by keeping the ratio $k=\kappa/\lambda$ fix and letting $\lambda \rightarrow 0$, while keeping $\mu = v_s \lambda /\sqrt{2}$ constant.  Since the only couplings between the Higgs doublets and the new Higgs singlet are dependent on $\lambda$ and $\kappa$, the singlet field decouples in this limit and one recovers the Higgs sector of MSSM. How this works will be explained in more detail in section \ref{sec:MSSMLimit}.  

Another thing worth mentioning about the NMSSM, is that it can be used to solve the little hierarchy problem of the MSSM. This is because if we let $\lambda$ (or $v_s$) become larger, the mass of the lightest Higgs gets larger, so by having a large $\lambda$ we can get a large Higgs mass. This approach is sometimes called $\lambda$-SUSY\cite{Barbieri:2006bg}. In that model we however give up the requirement of perturbativity up to the GUT-scale.

\section{Details of NMSSM} \label{NMSSMdetails}
In this section I will go through some technical details about the Higgs sector of the NMSSM. First the mass matrices are described in some detail, and then the couplings of the Higgses to the W/Z and the quarks are briefly described, introducing the concept of reduced couplings in order to easily compare it with the standard model Higgs and the MSSM. Finally some theoretical and experimental limits on the parameter space are discussed. Some other articles discussing the Higgs sector of the NMSSM are \cite{Mahmoudi:2010xp, Miller:2003ay}. 

\subsection{The mass matrices}

Since mixing only can occur between states with the same quantum numbers, we get three different mass matrices, one for the charged Higgs states, one for the scalar or CP-even neutral states, and one for the pseudoscalar or CP-odd states. Since they are obtained by taking derivatives of the potential, they are all real and symmetric. This is at tree level, taking higher order corrections into account this is no longer the case. The mixing matrices being real also means that, at tree level, there is no CP violation.
\subsubsection{Neutral scalar states}
In the natural basis $\{ H_{u,R}, H_{d,R}, S_R\}$ where the subscript $_R$ denotes the real part of the corresponding scalar field, we get the mass-squared matrix as follows
\begin{eqnarray}
M_{s,11}^2 &=&  \frac 1 4 g^2 v_u^2 + \frac{\lambda v_s v_d}{2v_u}(\sqrt{2}A_\lambda + \kappa v_s) \\
M_{s,22}^2 &=& \frac 1 4 g^2 v_d^2 + \frac{\lambda v_s v_u}{2v_d}(\sqrt{2}A_\lambda + \kappa v_s)\\
M_{s,33}^2 &=& v_s \kappa \left(\frac{1}{\sqrt{2}}A_\kappa + 2 v_s \kappa \right) + \frac{1}{\sqrt 2}\lambda A_\lambda \frac{v_d v_u }{v_s}\\
M_{s,12}^2 &=& v_d v_u (\lambda^2 -\frac 1 4 g^2) - \frac 1 2 v_s \lambda (\sqrt{2}A_\lambda + v_s \kappa) \\
M_{s,13}^2 &=& \lambda \left( v_s (v_u \lambda - v_d \kappa) - \frac{1}{\sqrt 2}A_\lambda v_d\right)\\
M_{s,23}^2 &=& \lambda \left( v_s ( v_d \lambda - v_u \kappa) - \frac{1}{\sqrt 2}A_\lambda v_u \right)
\end{eqnarray}
where I've used the stability conditions to eliminate the mass parameters from the potential in favour of the vacuum expectation values and couplings. This matrix doesn't really lend itself to much further algebraic simplification, so it is evaluated in the form given here and numerical methods are used to find it's eigenvalues, which corresponds to the physical masses of the CP-even Higgs states, which are denoted $H_1, H_2, H_3$, ordered from the lowest mass to the highest.

\subsubsection{CP-odd neutral states} 
In the natural basis $\{ H_{u,I}^0, H_{d,I}^0, S_I \}$ we get the following mass matrix for the pseudo-scalar states:
\begin{eqnarray}
M_{p,11}^2 &=& \frac 1 2 \frac{v_d }{v_u}v_s\lambda ( \sqrt{2}A_\lambda + v_s \kappa) \\
M_{p,22}^2 &=&\frac 1 2 \lambda v_s \frac{v_u}{v_d}  (\sqrt{2}A_\lambda + v_s\kappa )\\
M_{p,33}^2 &=& -\frac{3}{\sqrt 2}A_\kappa v_s \kappa + \frac{v_d v_u}{v_s}\lambda \left (\frac{1}{\sqrt 2}A_\lambda + 2\kappa v_s \right)\\
M_{p,12}^2 &=& \frac 1 2 \lambda v_s  (\sqrt{2}A_\lambda + \kappa v_s)\\
M_{p,13}^2 &=& \frac 1 2 v_d \lambda (\sqrt{2}A_\lambda - 2 v_s \kappa)\\
M_{p,23}^2 &=& \frac 1 2 v_u \lambda (\sqrt{2}A_\lambda - 2 v_s \kappa)
\end{eqnarray}
If the first two basis-elements are rotated with the angle $\beta$, a massless Goldstone mode decouples, and the new mass matrix (dropping the massless mode) in the basis $(P_1, P_2)$ becomes
\begin{eqnarray}
M_{p',11}^2 &=& \frac{v^2}{2v_u v_d}v_s \lambda (\sqrt{2}A_\lambda + \kappa v_s) \equiv M^2_A  \\
M_{p',22}^2 &=& \frac{v_u v_d}{v_s}\left( \frac{1}{\sqrt 2}A_\lambda + 2 v_s\kappa \right) - \frac{3}{\sqrt 2}A_\kappa v_s \kappa\\
M_{p',12}^2 &=& v\lambda(A_\lambda - 2\kappa v_s )  
\end{eqnarray}
where we introduce the mass parameter $M_A^2$. Note that this is not a physical mass, only a parameter which can be taken as one of the parameters instead of $A_\lambda$. It can be useful, because in the MSSM-limit, $M_A$ becomes the physical mass of the pseudoscalar Higgs. The matrix that diagonalises this is of course a $2\times 2$ orthogonal matrix, and can thus be parametrized by an angle $\theta_A$. The new basis in which the mass matrix is diagonal is then 
\begin{equation}
\left(\begin{array}{c}A_1\\A_2\end{array}\right) = 
\left( \begin{array}{cc} \cos \theta_A & \sin\theta_A \\
-\sin \theta_A & \cos \theta_A \end{array} \right)
\left(\begin{array}{c}P_1\\P_2\end{array}\right).
\end{equation}

\subsubsection{Charged states}
In the natural basis $\{ H_{u,R}^+, H_{d,R}^- \}$, the mass matrix for the charged states looks like
\begin{eqnarray}
M_{c,11}^2 &=&  \frac{1}{4}v_d^2 ( g_2^2- 2 \lambda^2) + \frac{v_d v_s \lambda}{2 v_u}\left ( \sqrt{2}A_\lambda + \kappa v_s \right )  \\
M_{c,12}^2 &=& \frac 1 4 v_d v_u (g_2^2 - 2\lambda^2) + \frac 1 2 \lambda v_s (\sqrt{2} A_\lambda + v_s \kappa) \\
M_{c,22}^2 &=& \frac 1 4 v_u^2(g_2^2/2- 2\lambda^2) + \frac{v_s v_u \lambda}{2v_d}(\sqrt{2}A_\lambda + v_s \kappa).
\end{eqnarray}
By a rotation through the mixing angle $\beta$, this gives a mass matrix in a new basis $\{ H^\pm, G^\pm \}$ with the only nonzero element
\begin{eqnarray}
M_{c',11}^2 &=& \frac{v^2}{2v_u v_d}v_s \lambda (\sqrt{2}A_\lambda + v_s \kappa) + \frac 1 4 g_2^2v^2 - \frac 1 2 \lambda^2v^2 \nonumber \\ 
&=& M_A^2 + M_W^2 - \frac 1 2 \lambda^2 v^2 
\end{eqnarray}
To get the second equality, we use the definition of $M_A^2$ in addition to the previously noted relation $(g_2v/2)^2 = M_W^2$. The state $G^\pm$ is a massless Goldstone mode, which is ``eaten'' by the $W^\pm$ to give it mass. The charged Higgs state is denoted by $H^\pm$.

\subsection{Reduced couplings}

If we want to express how the physical Higgs particles, i.e.\ the mass eigenstates, couple to fermions and gauge bosons, what one needs to do is to express the original weak eigenstates $H_u, H_d,S$ in terms of the mass eigenstates $H^\pm, A_1, A_2, H_1,H_2,H_3$. This is of course done by looking at the matrices that rotates the weak eigenstates into the mass eigenstates, i.e.\ the mixing matrices as defined above. 

We are primarily interested in how the $V$-boson couples to the different Higgses, where $V$ can be either $W^\pm$ or $Z$, and the couplings to quarks, since in the generic Higgs decay $H \rightarrow f\bar f$, there is a factor $\frac{m_f^2}{m_W^2}$ meaning that the heaviest fermion allowed dominates, i.e.\ either the top or bottom quark. 

The way to find these couplings is to write down the relevant terms in the Lagrangian, which is originally in terms of the weak eigenstates, and then re-express it in terms of the mass eigenstates $H_i, H^\pm$ and $A_j$. The details can be found in \cite{Miller:2003ay}.

In order to simplify the notation and keep it from getting unneedingly cluttered, we define so-called \emph{reduced couplings}, where we take the full coupling and divide out the associated SM coupling,
\[ G_{VVH_i} \equiv \frac{g_{VVH_i}}{g^{SM}_{VVH}}, \hspace{4 mm} G_{ZA_iH_j} = \frac{g_{ZA_i H_j}}{G^{SM}_{ZHH}} = \frac {g_{ZA_i H_j}}{g/2}, \]
where $g = \sqrt{g_1^2 + g_2^2}$, $g_1$ and $g_2$ being the gauge couplings of the electroweak force. Here, $V$ can stand for either W or Z, the reduced coupling will be the same in either case. The reduced coupling we will look the most at is the $H_i VV$ coupling, since this measueres how standard model like the scalar Higgses are. If we let $S_i = (H_{u,R}, H_{d,R},S_R)$ be the scalar weak eigenstates, and $H_i = \sum_j S_{ij}S_j,$ (i.e.\ $S_{ij}$ is the mixing matrix), then this reduced coupling is defined as\cite{Mahmoudi:2010xp} 
\begin{equation}
G_{H_iVV} = \sin \beta S_{i1} + \cos \beta S_{i2}.
\end{equation}
The $H_i H^\pm W^\mp$ coupling is similarly given by
\begin{equation}
G_{H_i H^\pm W^\mp} = \cos \beta S_{i1} - \sin \beta S_{i2}.
\end{equation}

Since these reduced couplings comes directly from orthogonal mixing matrices, we may conclude that they should fulfil certain sum rules. This is because of the sum rules that elements of orthogonal matrices fulfil: the sum of the squares of one row (or column) is equal to one. For the reduced couplings, this means
\begin{equation}
\sum_i G^2_{ZZH_i} = 1, \hspace{4 mm} \sum_{i} G_{H_i H^\pm W^\mp}^2 = 1.
\end{equation}
In the same way, the reduced couplings of the Higgses to the top and bottom quarks also comes directly from an orthogonal matrix, but in this case a dependence on $\tan \beta$ also enters, since this describes how large the difference is between the two VEVs of the Higgs doublets. In this case the sum rules are
\begin{equation}
\sum_i G_{H_i tt}^2 = \frac{1}{\sin^2 \beta}, \hspace{4 mm} \sum_i G_{H_i bb}^2=\frac{1}{\cos^2 \beta}.
\end{equation}
We also have the sumrules from the columns, for example
\begin{equation}
G^2_{H_iVV} + G_{H_i H^\pm W^\mp}^2 + S_{i3}^2 = 1, \hspace{ 3 mm} i=1,2,3 
\end{equation}
where $S_{i3}$ is the singlet component of $H_i$. If $S_{i3}\approx 0$, then $H_i$ will be purely doublet and the corresponding sum rule $G^2_{H_iVV}+G^2_{H_i H^\pm W^\mp} = 1$ is recovered. Conversely, if $S_{i3} \approx 1$ then both the other couplings will be suppressed, which means that detection of $H_i$ will be difficult. These sum rules are quite trivial in nature, but can be a useful check on the numerical methods used. They are also important phenomenologically, since they in effect is a good measure how standard model like the different $H_i, A_i$ are. For example, in this last sum rule, if the first term is large it means that $H_i$ is SM like, if the second term is large it means there's a large coupling between the doublets making $H_i$ MSSM like, and the last term corresponds to how singlet-like $H_i$ is.

Of course, since the standard model only has one (scalar) Higgs particle, only the $H_1$ couplings (assuming that the lightest Higgs also will be the standard model like) have a direct correspondence in the standard model. Nevertheless we can define reduced couplings by scaling away the gauge couplings and masses.

\subsection{Constraints on the parameters}
In this section I will explain some theoretical and experimental limits on the parameter space, and motivate the choices of parameters later used when studying some numerical results. I will discuss for which intervals it is sensible to choose values for the parameters, which I choose as $\lambda, m_{H^\pm}, \kappa, \tan \beta$ and $A_\kappa$. Some limits can be found from theoretical considerations and requirements, while others come from experiments at accelerators or astrophysics. 

As said above, I use the value $v=246$ GeV for the electroweak scale, and choose the `standard' value $\mu = \lambda v_s/\sqrt{2} = 200$ GeV, which allows me to see $v_s$ as a fixed value when $\lambda$ is chosen. There is a restriction from LEP \cite{Schael:2006cr} on the minimal size in $\mu$, requiring that $|\mu| > 100$ GeV, coming from lower limits on Higgsino masses, but this is really for MSSM. Nevertheless, a too small value of $|\mu|$ doesn't work. 

Since my analysis is at tree level, we will not discuss parameters entering at loop level, where the Higgs masses get corrections depending on for example the top and the stop masses. There is probably many cases where even at tree level limits from measurements could be used to rule out large parts of the parameter space, but doing this in detail is regretfully beyond the limited scope of this study.

A general way to restrict the parameter space is to require that they fulfil some grand unified scenario where all couplings of the same type gets the same value at the GUT scale. This is called universal boundary conditions, and will not be required here.

\subsubsection{$\lambda$ and $\kappa$}
We can see that with $\kappa = 0$, the Lagrangian, (\ref{NMSSM-Lagrangian}), has an additional $U(1)$ symmetry, called \emph{Peccei-Quinn} symmetry\cite{PQ} (henceforth called PQ-symmetry).  This symmetry was proposed as a solution to the strong CP problem, i.e.\ the problem of explaining why QCD doesn't seem to violate CP symmetry like the electroweak interactions do. If this symmetry is exact, i.e.\ $\kappa =0$, it will be spontaneously broken by the nonzero VEV of the singlet scalar, which will give rise to a massless Goldstone boson, called the Peccei-Quinn \emph{axion}. This axion will show up as the extra pseudoscalar Higgs field (compared to in the MSSM). However, this case can in principle be ruled out since it would since there are lower bounds on allowed axionmasses\cite{ParticleDataGroup} which only can be avoided if $10^{-16} <\lambda <10^{-7}$\cite{Miller:2003ay}. Such a small value of $\lambda$ would mean that $v_s$ would have to grow very large, making the model unattractive as a solution to the $\mu$-problem.

So from this we conclude that we need a nonzero value of $\kappa$, breaking the PQ symmetry. The size of the $\kappa$ coupling will regulate how badly this symmetry is broken, and with a small value, only slightly breaking PQ symmetry, we will get a nonzero mass for the lightest pseudoscalar. 

If one uses the requirement that $\lambda, \kappa$ and the Yukawa couplings should stay small (eg. $< 1$) so that perturbation theory can be used up to the GUT scale, and uses the renormalization group flow, one can get the approximate limit at the electroweak scale\cite{Miller:2003ay} 
\begin{equation}
\sqrt{\lambda^2 + \kappa^2} \lesssim 0.7
\end{equation}
Also, from choosing a large number of different values of $\lambda$ and $\kappa$ at the GUT scale and using the renormalization group equations to run them down to the electroweak scale, one can see that the flow favours a small $\kappa$ value. 

We also note that if $\lambda$ gets too small, this forces $v_s$ to become big, which means that the model no longer works well as a solution to the $\mu$-problem. Even if we allow $v_s$ to take a value of a few TeV, say 2 TeV, which is well over but still ``close'' to the electroweak scale in some sense, this places a limit on $\lambda \gtrsim 0.1$, so we get a rather stringent condition on $\lambda$. 

If universal boundary conditions at the GUT scale are imposed (which gives us the so called constrained NMSSM\cite{Djouadi:2008uj}), we also get that the ratio $\lambda/\kappa$ has to be close to 3. This is however not something that will be exclusively used since I don't in general impose universal boundary conditions.  

\subsubsection{$\tan \beta$ and $A_\kappa$ }

\begin{figure}

	\centering
	\includegraphics[width= 70 mm, angle=270]{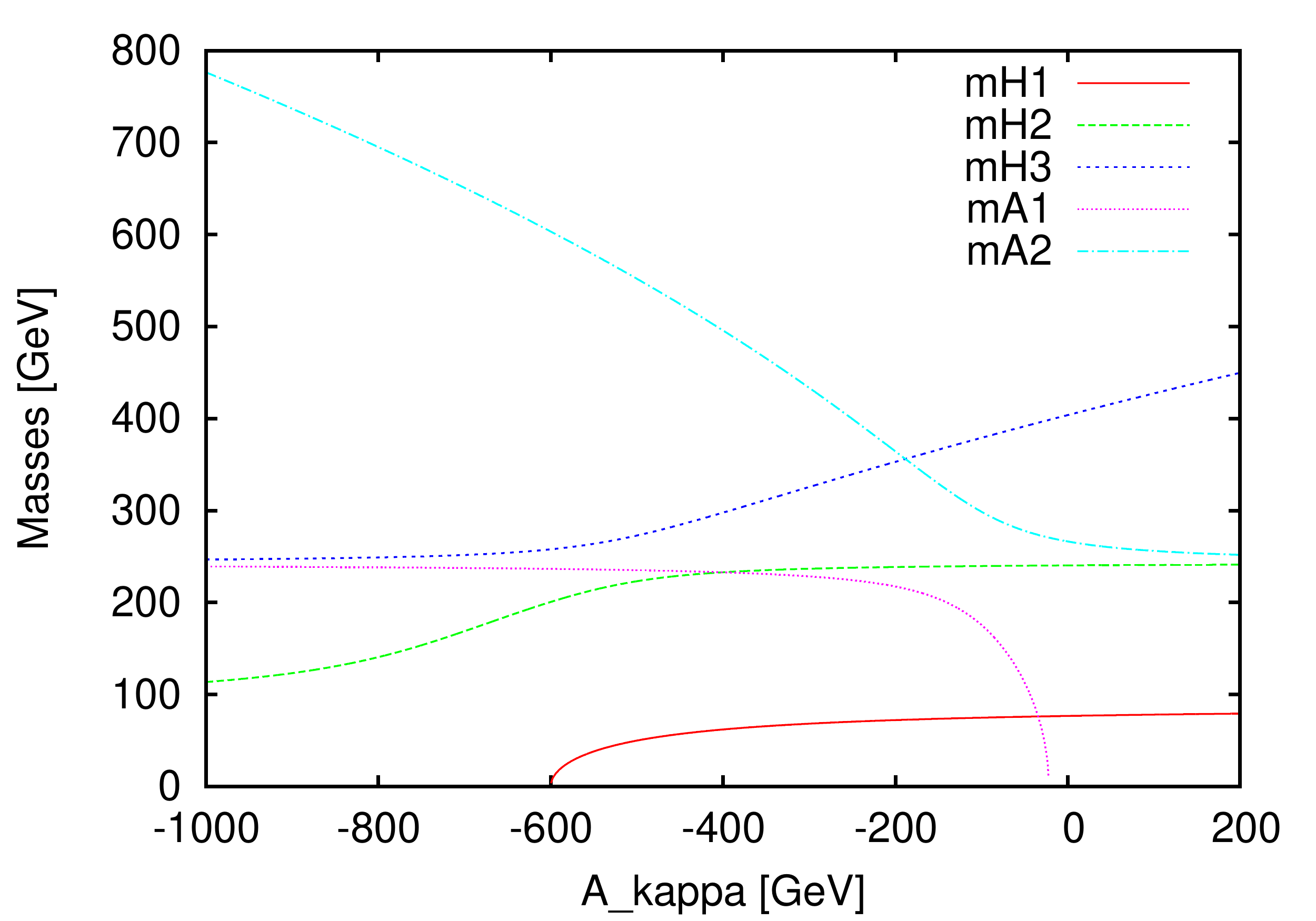}

\caption{The masses as a function of $A_\kappa$, where $\tan \beta = 7, m_{H^\pm} = 250 \mbox{ GeV}, \kappa = 0.3$ and $\lambda = 0.3 $. }
\label{Akappaconstraints}
\end{figure}

The range of $A_\kappa$ is rather tightly constrained from the condition of vacuum stability. In figure \ref{Akappaconstraints}, the only allowed range of $A_\kappa$ is where all the masses are positive, i.e.\ $-600 \lesssim A_\kappa \lesssim -30$ GeV. In many cases the limits are a lot stricter than this. From such plots you can also see that for some choices of the other parameters there are no acceptable value of $A_\kappa$ at all; in some cases the lightest pseudoscalar and the lightest scalar never both get a positive mass at the same time. For the coming plots where the $A_\kappa$ dependence matters, the value of $A_\kappa$ is chosen roughly in the middle of its allowed range, for a typical value of the running parameter. The typical values are $-100$ and $-250$ GeV.

As for $\tan \beta$, an analysis of the running couplings shows that a low value of $\tan \beta$ is favoured. However, experiments rule out a too small value, so a not so small value is required \cite{Mahmoudi:2010xp}. 

The theoretical upper bound is $\tan \beta \lesssim m_t / m_b \sim 50$, and I will briefly study what happens when you take a large value, $\tan \beta = 30$ in the model. See also \cite{Ananthanarayan:1995zr} for a study of what happens when you saturate this upper bound. We also have a lower bound $\tan \beta \gtrsim 1.2$ from requiring $\lambda_t$, the top quark Yukawa coupling, to remain small up to the GUT-scale\cite{Mahmoudi:2010xp}.

\subsubsection{$m_{H^\pm}$ and the other Higgs masses}
For the Higgs masses experiments have placed general lower limits. For the neutral Higgses, LEP has published negative search results\cite{Schael:2006cr} in some different decay channels, and depending on the precise branching ratios the limits looks a little bit different, but generally the lower bound from LEP is around $m_{H_i} > 80-90$ GeV, for the MSSM. In the NMSSM these limits can be avoided, but they still give some kind of general guidelines. In the numerical studies, a lower value of $m_{H^\pm} = 90 $ GeV, and a higher value of $250$ GeV will be used, when we don't let $m_{H^\pm}$ vary.

From different experiments at the Tevatron and LEP, rather strict limits can be placed on a SM-like Higgs\cite{ParticleDataGroup}, for example the Tevatron has excluded the range $158\mbox{ GeV}<m_h<173\mbox{ GeV}$, and from LEP we have exclusions of a mass lower than 114 GeV, but since the couplings to fermions and gauge bosons of the MSSM or NMSSM Higgs bosons can be suppressed compared to the SM Higgs, these limits can be avoided. 
\section{Results}
In this section I will present numerical results that explores some of the features of NMSSMs parameter space. First, masses and couplings are treated as a function of the charged Higgs mass. From this we see some possibly interesting features of the model. Then it is studied how the masses and reduced couplings (and thus the mixing) behave in some different kinds of MSSM limits.

Since my calculation is only at tree level, it is not sensible to compare directly with experimental limits. Even so, the general features are maybe even better understood at tree level, since it is easier to compare directly with the formulae without too much cluttering of the expressions. For my numerical results, I've written code in Java, using the basic linear algebra library \textsc{Jama} to diagonalize and find eigenvalues of matrices.

\subsection{Varying the charged Higgs mass}
In order to see how the NMSSM mass spectra behaves, it can be instructive to plot the masses as a function of the charged Higgs mass. From these plots, and the requirement of vacuum stability (i.e.\ $m^2_{H_1} > 0$, the lightest scalar mass positive) we can find limits on allowed values for $m_{H^\pm}$ for fixed values of the other parameters.

As an aside, just in order to confirm the theory and my numeric calculation, we can check that the sum rules for the reduced couplings actually holds in practise, which it turns out they do. A thing to note when looking at plots of reduced couplings and $\cos \theta_A$ is that what I really plot is the absolute value of the couplings. This is for two reasons, first of all that all we really care about is the strength of the different couplings, the sign can of course matter (mostly when you go to higher orders) but not for our purposes here. The other reason is that the numerical method used switched signs discontinuously, so without taking the absolute value the graphs looks very discontinuous and strange. This can probably be fixed rather easily, but since the sign doesn't matter for our purpose no effort was expended on this.

\begin{figure}

	\centering
	\subfloat[Masses]{\includegraphics[width= 60 mm, angle=270]{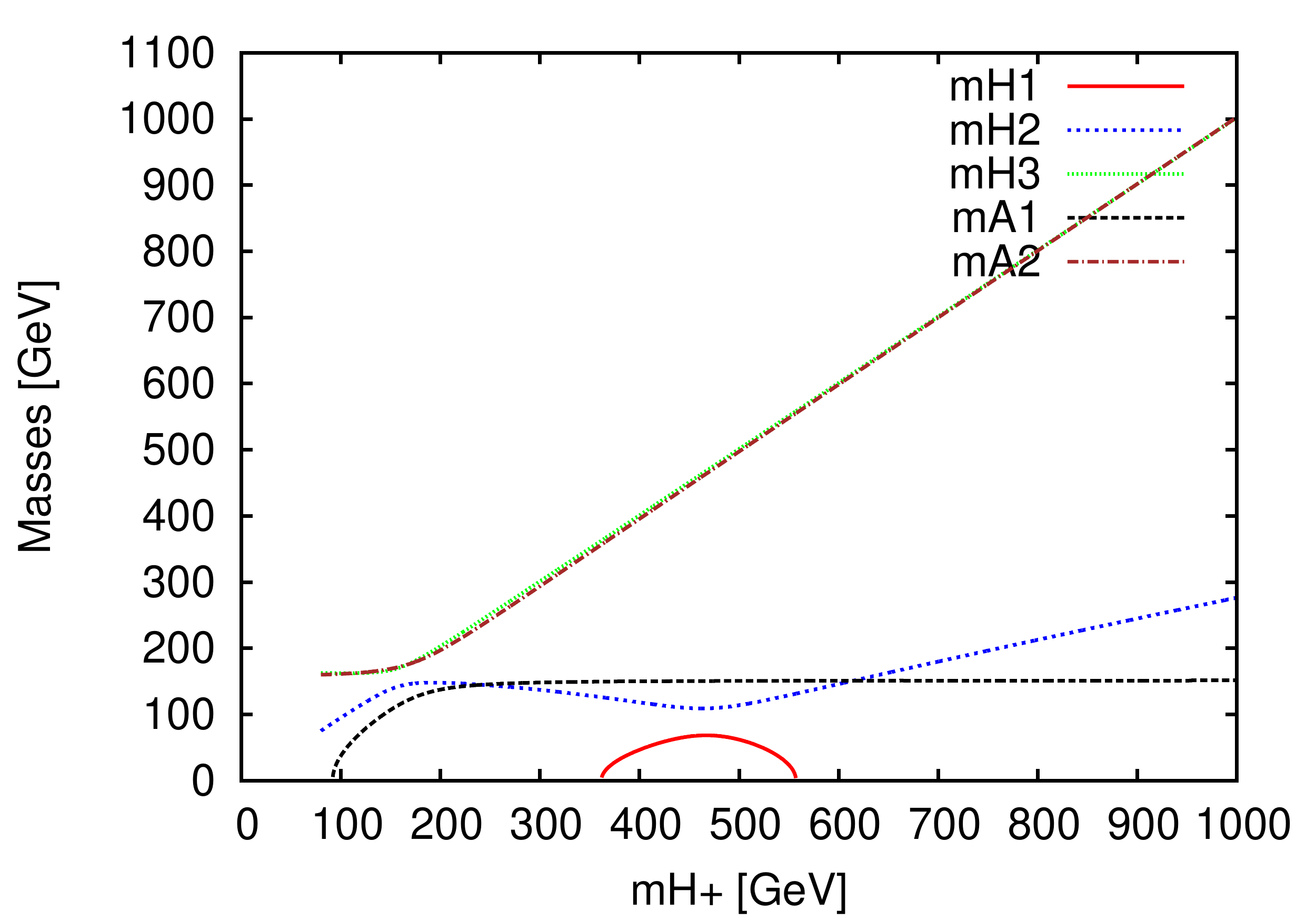}}
\subfloat[$H_iVV$-couplings and $\cos \theta_A$]{\includegraphics[width=  6 cm, angle=270] {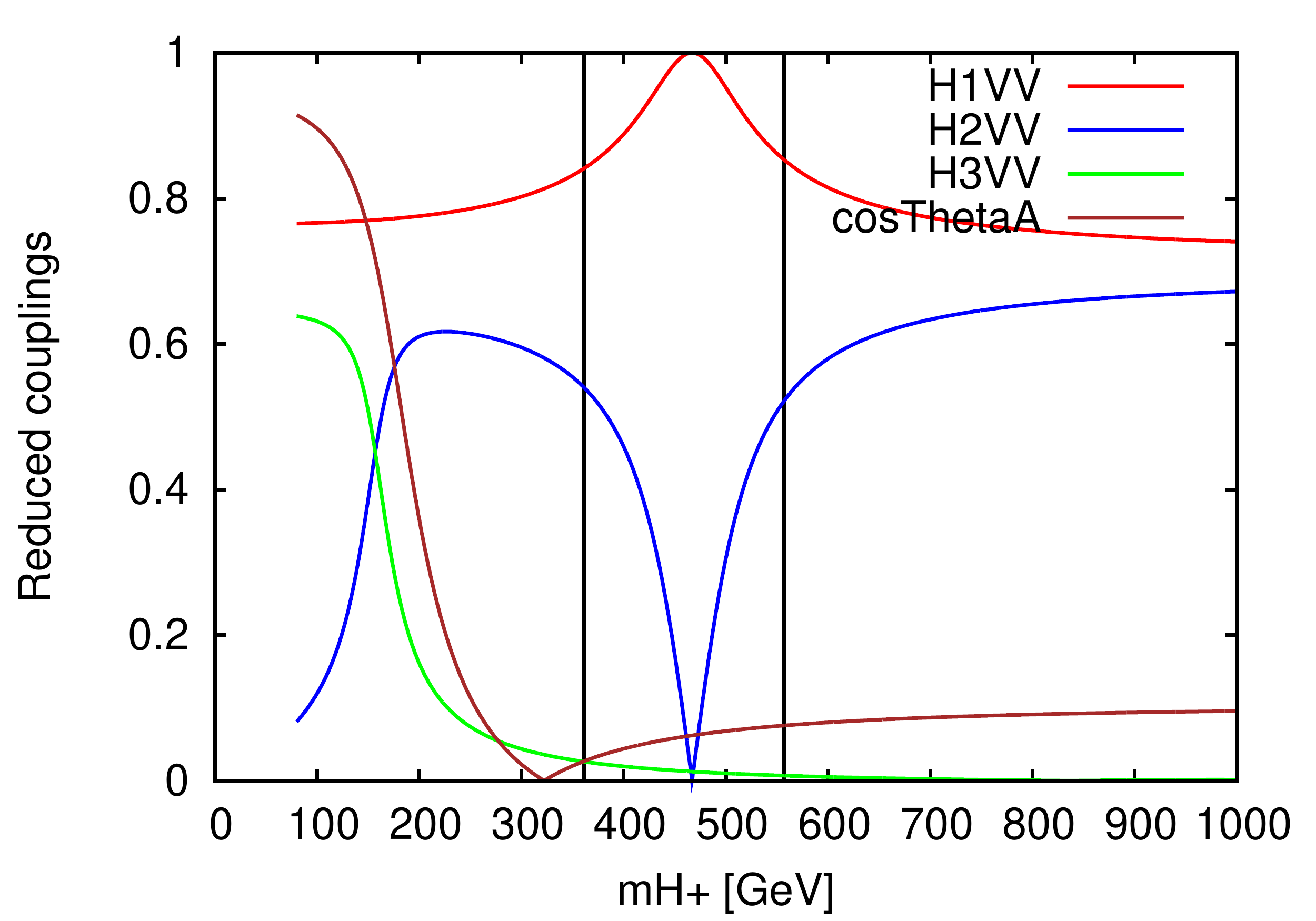}} \\
\subfloat[$H_iH^\pm W^\mp$-couplings] {\includegraphics[width= 6 cm, angle=270] {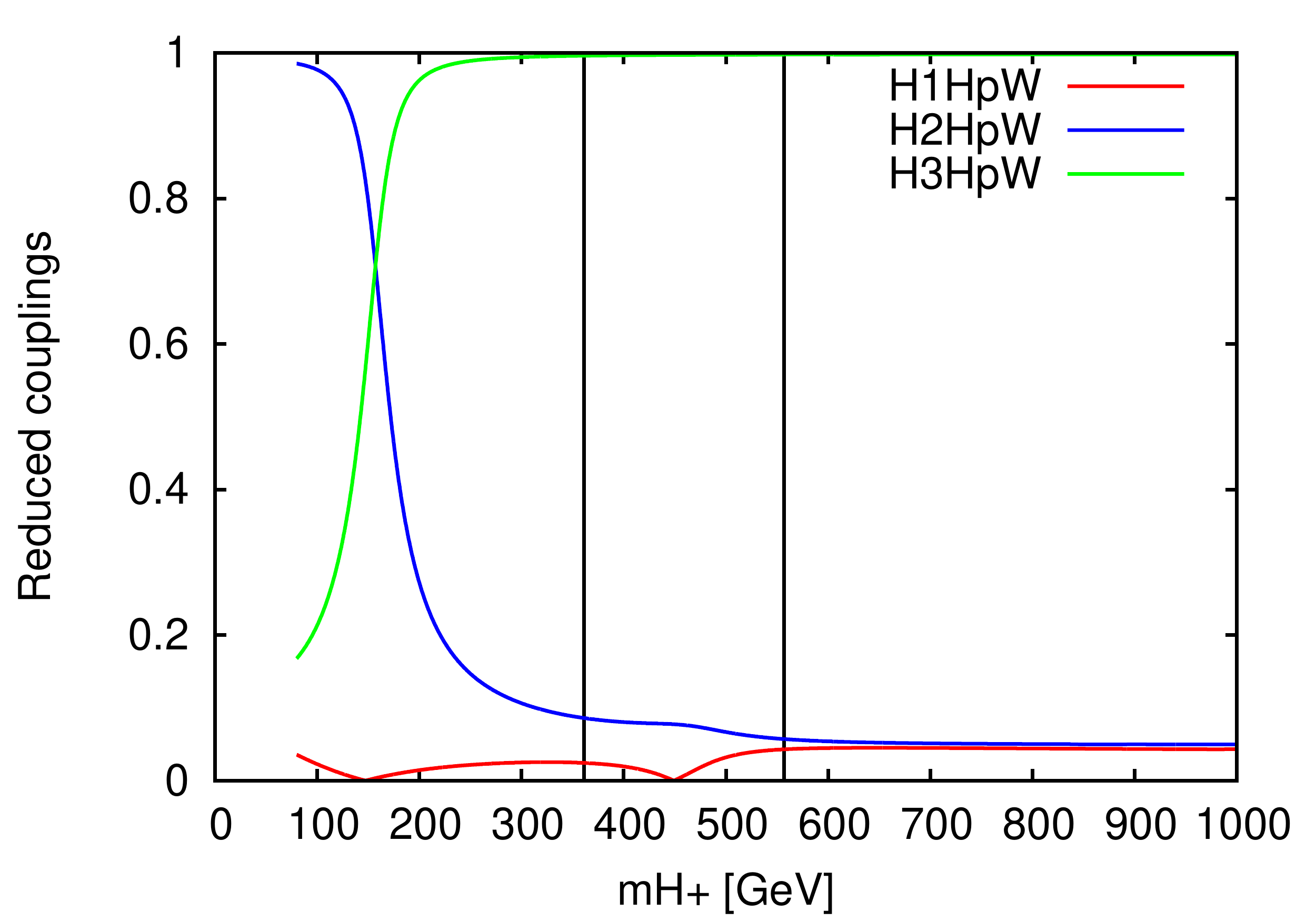}}
\subfloat[$S_{i3}$] {\includegraphics[width= 6 cm, angle=270] {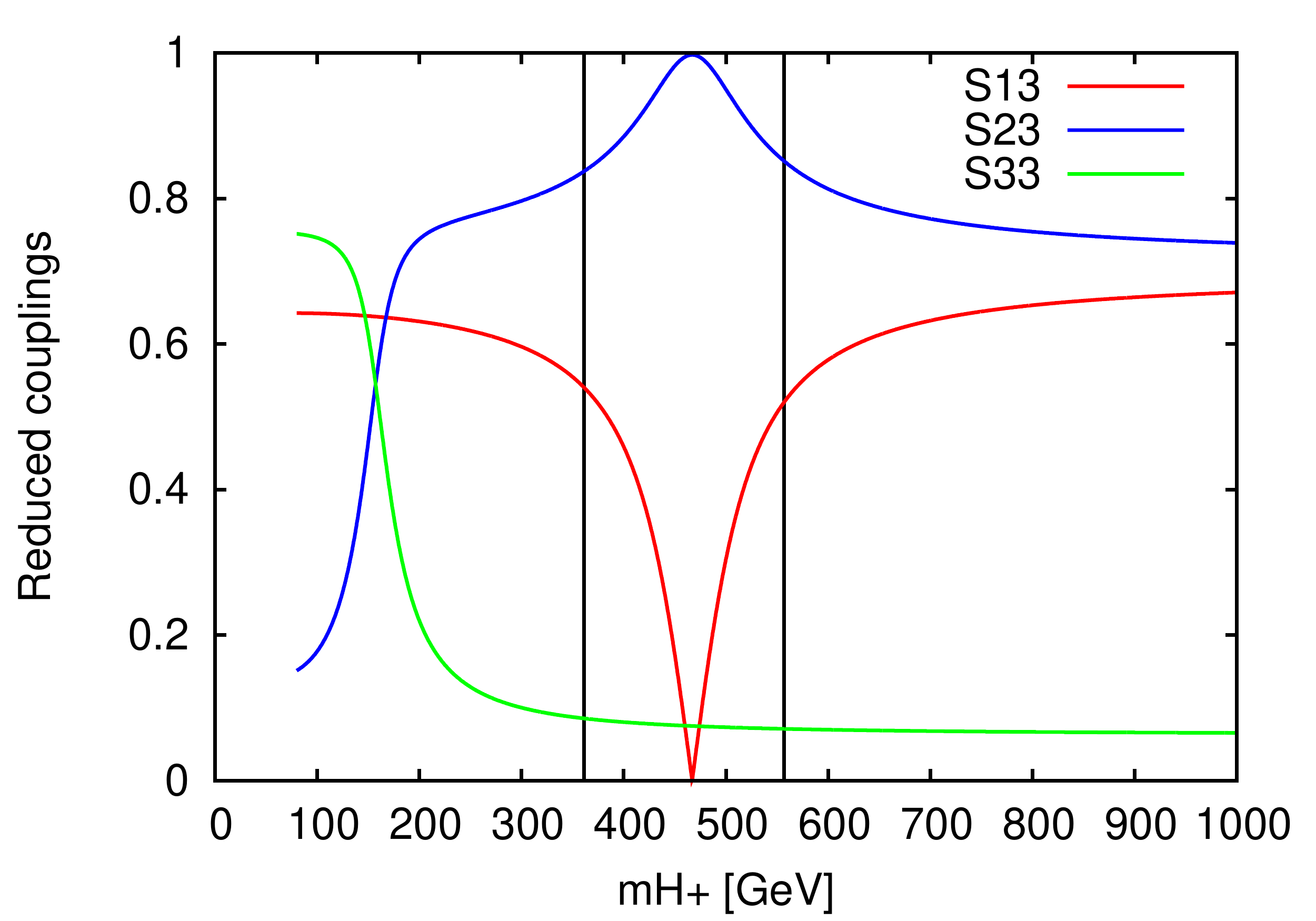}}
\caption{The Higgs masses, some different scalar couplings and $\cos \theta_A$ as functions of $m_{H^\pm}$, for $\kappa = 0.1, \lambda = 0.3, A_\kappa = -100$ and $\tan \beta = 2$. In (b,c,d), the physical range of the parameter space is inside the two black lines. $S_{i3}$ is the singlet component of the $H_i$ state.}
\label{mCharged1}
\end{figure}
\subsubsection{The NMSSM with a small $\kappa$}
In figure \ref{mCharged1} we can see that for the following choices of parameters, $\lambda = 0.3$, $\kappa = 0.1$, $\tan \beta = 2$, $A_\kappa = -100$ GeV, $m_{H^\pm}$ has to be between 360 and 550 GeV. This case is representative for a small $\kappa$-value, which corresponds to a slightly broken PQ-symmetry. This is favoured by the renormalisation group flow. From the figure we also see that the mass of the lightest pseudoscalar almost doesn't change as soon as $m_{H^\pm}$ gets above $\sim 250$ GeV. This is the singlet dominated pseudoscalar whose mass mainly comes from the $\frac 1 3 \kappa S^3$ term in the Lagrangian. The heavier states that grows more or less linearly are the fields with little mixing with the singlet, (as seen from that $S_{33}$ and $\cos \theta_A$ both are small) and behave the same way as in the MSSM. For these parameters, the next lightest scalar state, $m_{H_2}$ also doesn't depend very strongly on $m_{H^\pm}$, and grows much slower than in MSSM.

So we see that even though the spectra of high mass states stay roughly the same, the three light Higgs states means that a NMSSM with parameters close to these will be easily distinguished from MSSM, even if we only find the lighter Higgses. Of course, this only works if the reduced couplings, $G_{H_i VV}$ and $G_{A_1H_iZ}$ doesn't become too small to prevent detection, which happens in a small part of the relevant parameter space.

In figure \ref{mCharged1}.b the couplings of the scalar fields to the W/Z bosons as well as $\cos \theta_A$ are plotted as a function of the charged Higgs mass. Since the reduced couplings can be thought of as a measure of the mixing between weak eigenstates, this plot shows that the mixing depends on the charged Higgs mass in a slightly complicated way. The lightest Higgs is as one might guess the most standard model like, and it is the mixing with $H_2$ which raises $m^2_{H_1}$ above zero. For a specific value, $m_{H^\pm} = 466$ GeV, the coupling $G_{H_2 VV}= 0$, so if this specific scenario is true, the next-lightest Higgs would be totally singlet-like (as we see in figure \ref{mCharged1}.d). In this case, the $H_2$ Higgs would be totally undetectable through the channels used to look for the standard model Higgs. We also see that in the physical range at least, the pseudoscalar mixing only varies a little. The $H_iH^\pm W^\mp$ couplings, which measure how doublet or MSSM-like the scalars are, vary very little in the physical range, but we do see that $H_1H^\pm W^\mp$ pass through zero when $m_{H^\pm} = 449$ GeV. We also see that the heavy $H_3$ is doublet-dominated.

From the figure we can also see that $m_{A_1}$ and $m_{A_2}$ seems to switch behaviour with respect to the charged Higgs mass around $m_{H^\pm}\sim 150$ GeV. This switch is also apparent in how $\cos \theta_A$ behaves. After this however, the pseudoscalar mixing stays more or less constant, and doesn't vary rapidly in the physical region as the scalar mixing does. In the same way we see the switch in behaviour between $m_{H_2}$ and $m_{H_3}$ reflected in how all of the corresponding couplings switch, although this happens outside the physical region.

\begin{figure}

	\centering
	\subfloat[Masses]{\includegraphics[width= 6 cm, angle=270]{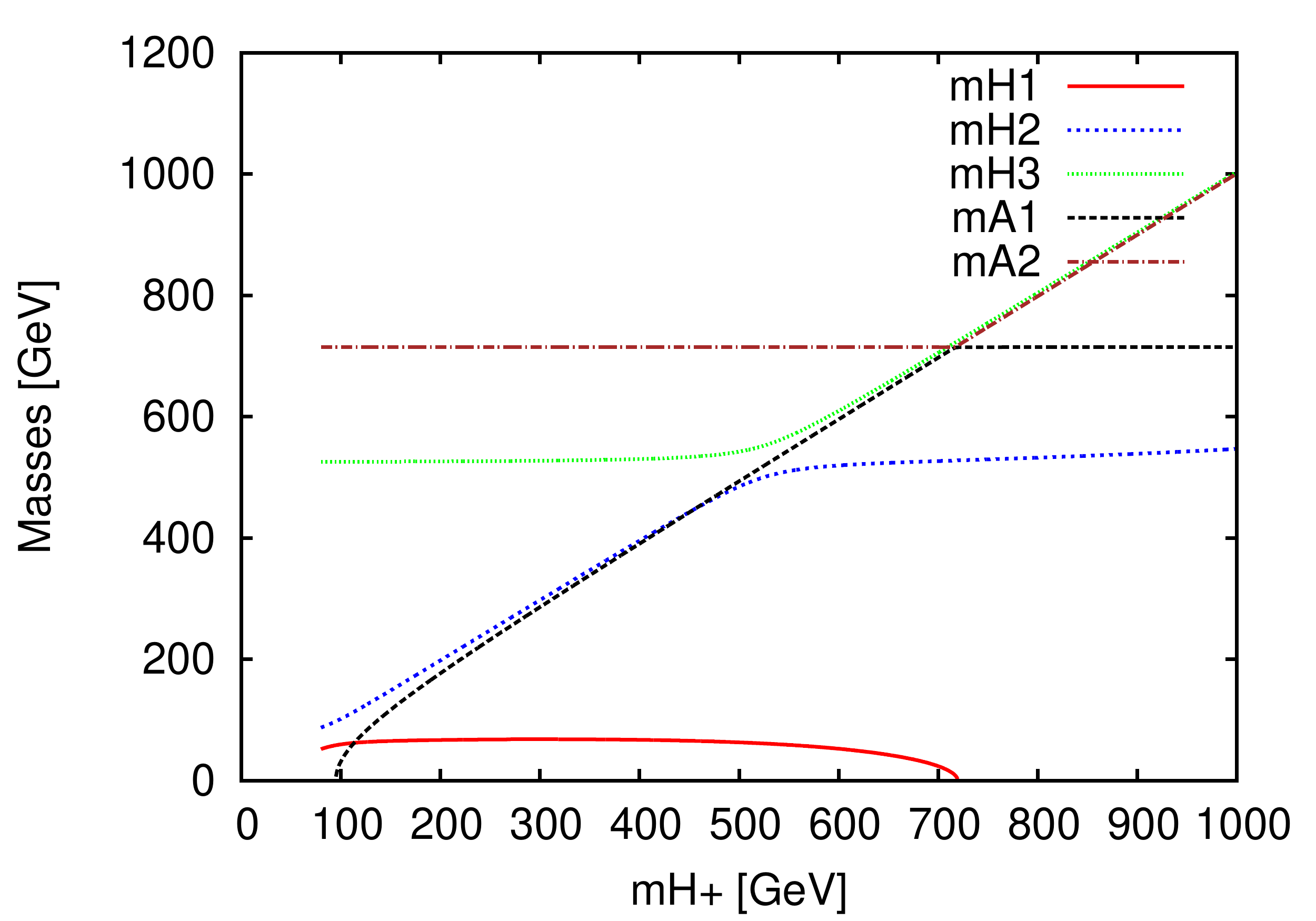}}
	\subfloat[$H_iVV$-couplings] {\includegraphics[width= 6 cm, angle=270] {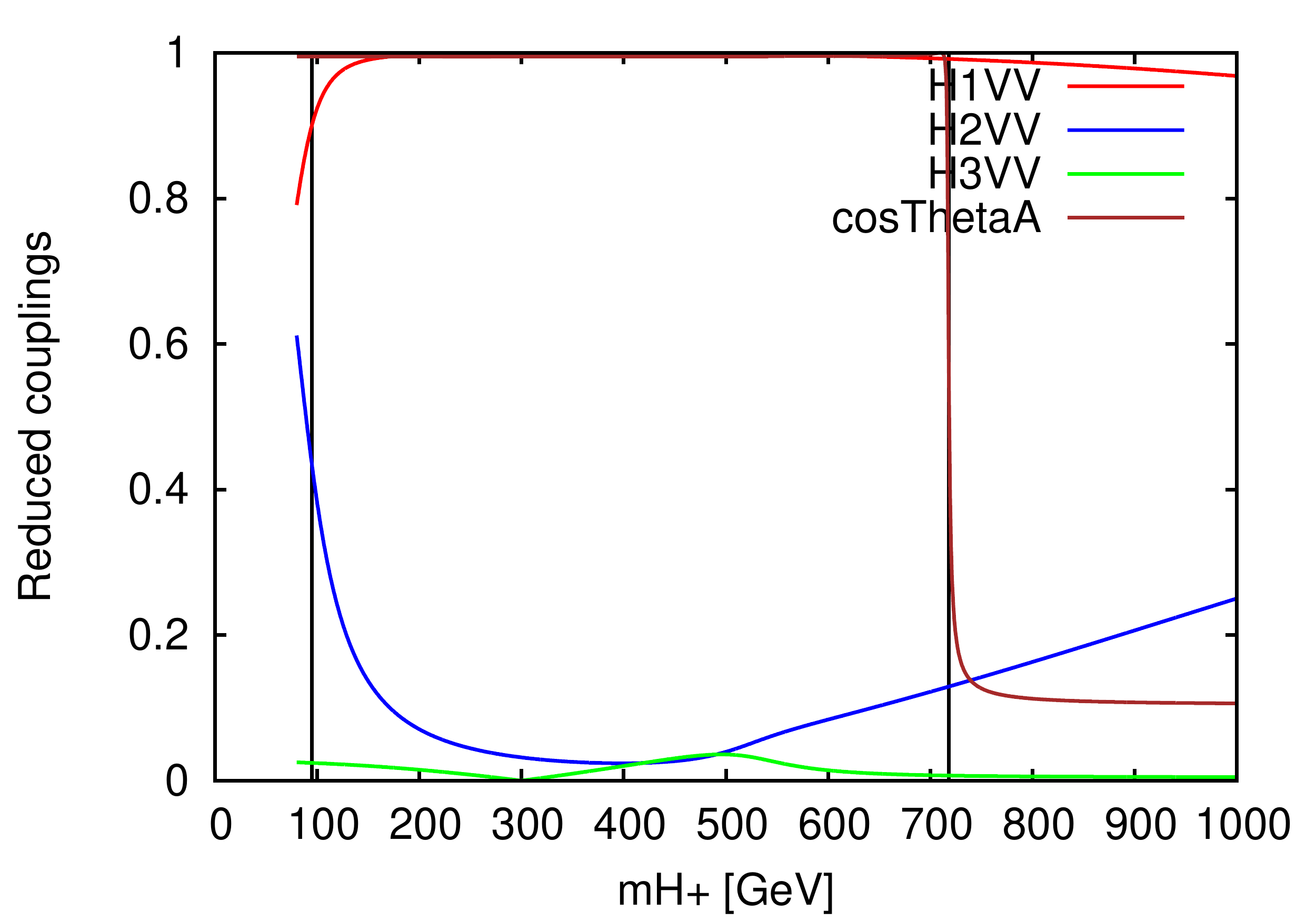}}\\	
	\subfloat[$H_i H^\pm W^\mp$-couplings] {\includegraphics[width= 6 cm, angle=270] {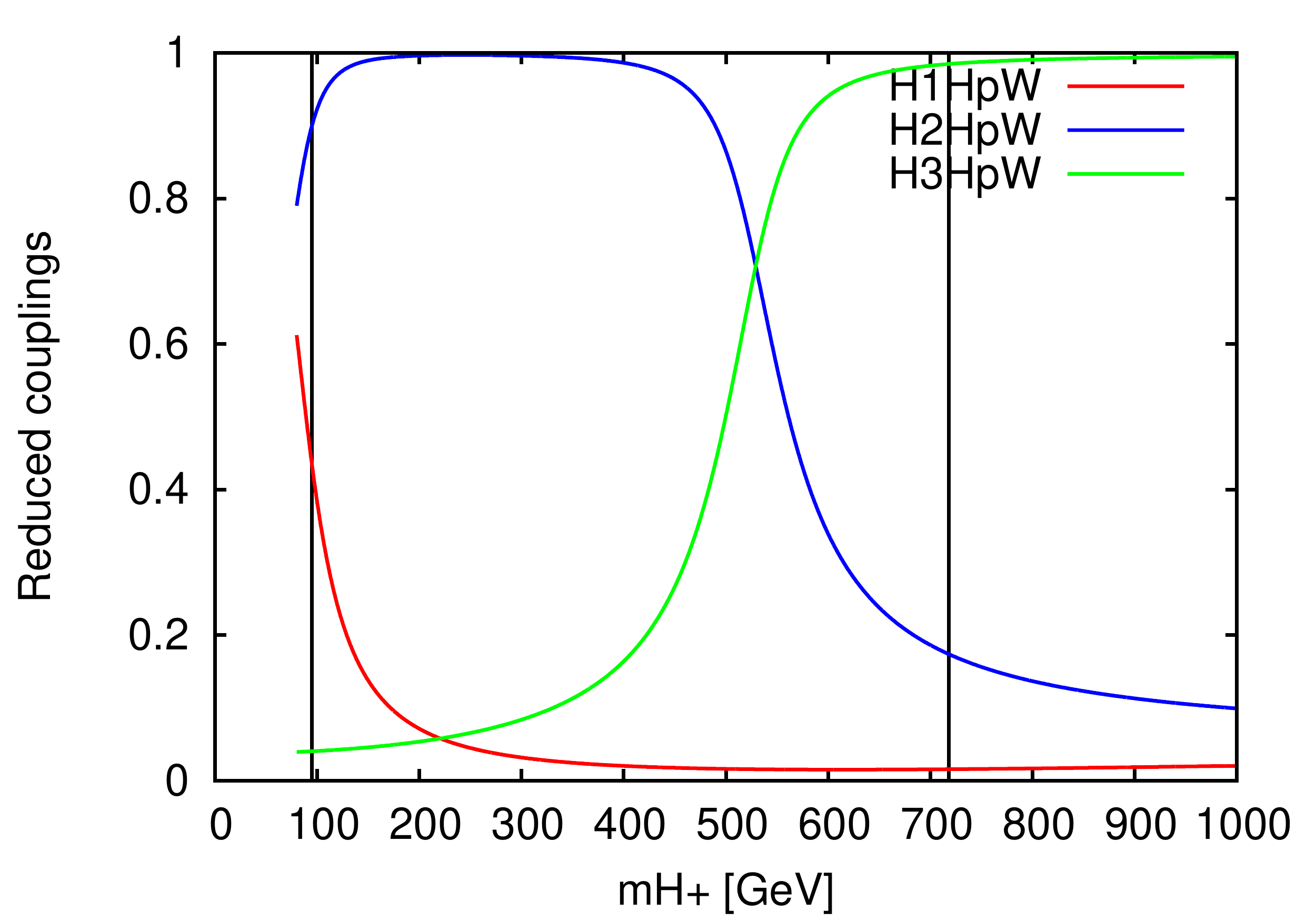}}
\subfloat[$S_{i3}$]{\includegraphics[width= 6 cm, angle=270]{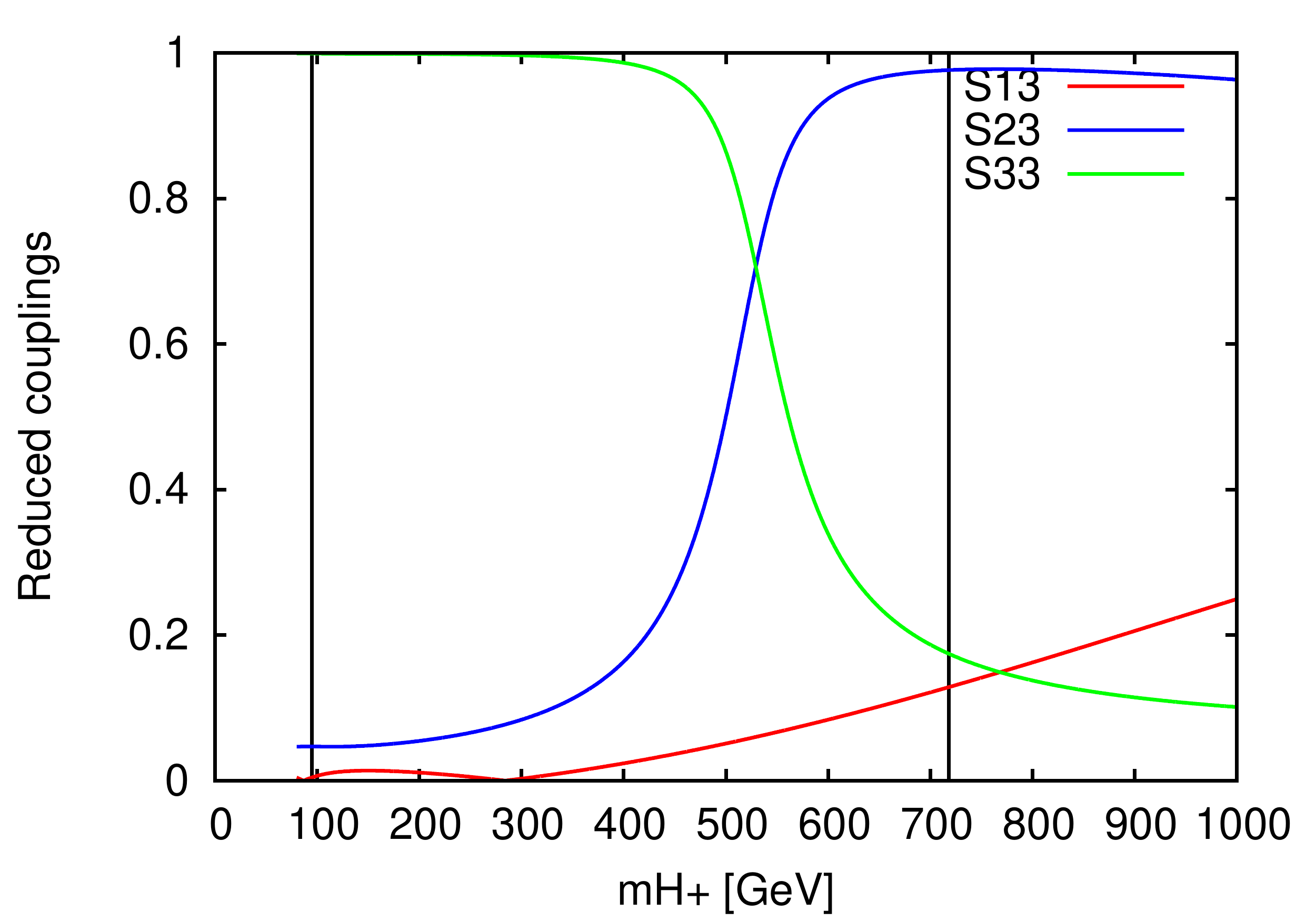}}

\caption{The Higgs masses, some of the scalar couplings and $\cos \theta_A$, as functions of $m_{H^\pm}$, for $\kappa = 0.5, \lambda = 0.3, A_\kappa = -500$ and $\tan \beta = 2$. In (b,c,d) the physical range is again between the black lines.}
\label{mCharged-high-kappa}
\end{figure}

\subsubsection{Larger $\kappa$}
If we let the value of $\kappa$ get larger, the PQ symmetry is more badly broken and the lighter pseudoscalar gets a larger mass. This is not favoured by the renormalization group flow, but we have no a priori reason to exclude it. In figure \ref{mCharged-high-kappa} we have plotted the mass spectrum and couplings as functions of $m_{H^\pm}$ for $\kappa = 0.5$. In this case, the lightest Higgs is the most standard model like by far, and $H_2$, $H_3$ again switch behaviour, around $m_{H^\pm} \simeq 520$ GeV. 

We also see that this large $\kappa$ loosens the constraints on $m_{H^\pm}$ from vacuum stability. The value of $A_\kappa = -500$ GeV used in the figure has been chosen approximately in the middle of its allowed range for these parameters. 

In this case, compared to the previous case with $\kappa$ small, we see that apart from the lightest Higgs, the rest of the masses are significantly larger. However, they are not extremely heavy and are still very much within the range of detection, but the spectrum of light Higgses present in the previous case is absent. This would make it harder in this case to distinguish between the MSSM and the NMSSM if only the two lightest Higgses can be detected, compared to the case with smaller $\kappa$.

Just as in the previous case, the pseudoscalar mixing, i.e.\ $\cos \theta_A$, stays rather constant except at the place where the two pseudoscalar fields switch identity. In this case this switching behaviour is more distinct, something we see both in how the masses and how $\cos \theta_A$ behaves.

\begin{figure}
	\centering
  \subfloat[Masses]{\includegraphics[width= 6 cm, angle=270]{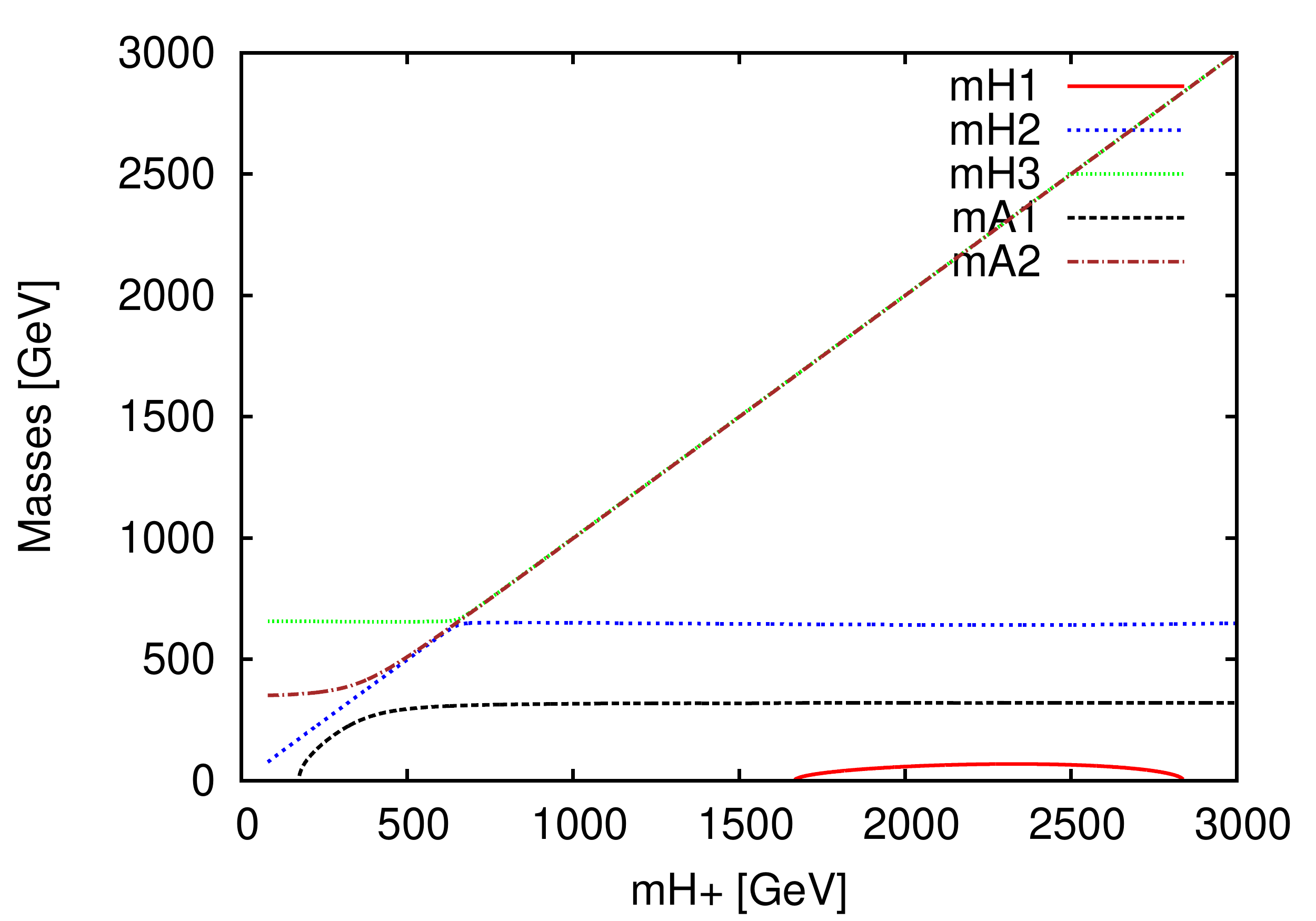}}
  \subfloat[$H_iVV$-couplings] {\includegraphics[width= 6 cm, angle=270] {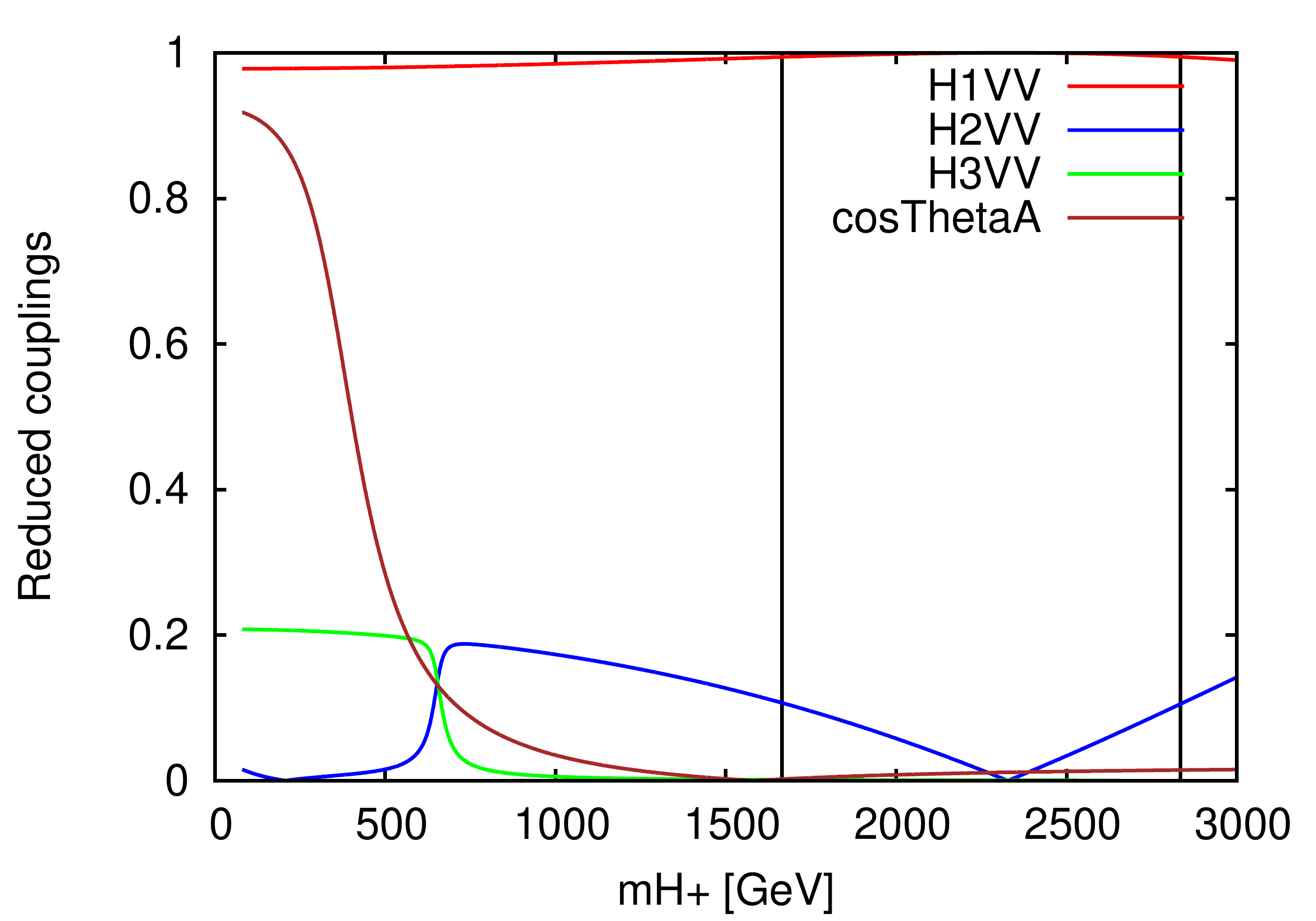}}

\caption{The Higgs masses and reduced couplings of the scalar states to the Z and W bosons, as functions of $m_{H^\pm}$, for a higher value of $v_s$, equivalent to $\mu = 1000$ GeV, instead of the normally adopted value of $\mu = 200$ GeV. The rest of the parameters are the same as in figure \ref{mCharged1}. In (b), the physical range is inside the vertical black lines.}
\label{high-mu}
\end{figure}

\subsubsection{A larger $v_s$ or $\mu$-value}
If we vary the expectation value of the singlet field, $v_s$, or equivalently the value of the effective $\mu = v_s \lambda$ parameter, this doesn't change the qualitative behaviour of the mass spectrum very much, but it changes the quantitative behaviour. All but the lightest Higgs gets heavier, including the charged Higgs, since the region of vacuum stability gets pushed upwards, see figure \ref{high-mu}. We also see that the constraint from vacuum stability is relaxed (see for comparison figure \ref{mCharged1}), and that the charged Higgs mass, as well as the masses of $H_3$ and $A_2$ are now forced to be larger than $\sim 1.5$ TeV. Since the lightest of the Higgses remains light, it means that a higher $v_s$ must make the $H_1$ more SM-like, which means that the $H_2$ becomes more singlet-like. That $H_1$ becomes doublet-like means that its coupling to W and Z should become large, and this is indeed also the case, as we can see in the right panel of figure \ref{high-mu}.

If we make the same plots for a large value of $\kappa$, the effect on the mass spectrum will be bigger and the heavy states will thus be even heavier, since it gets amplified by the large $\kappa$ value. In this case the heavy singlet dominated fields will decouple and the lower mass states will behave like in the MSSM, so in this case the distinction between NMSSM and MSSM will be hard to find. 

\subsubsection{Large $\tan \beta$ values}

\begin{figure}
	\centering
	\subfloat[Small couplings]{\includegraphics[width= 60 mm, angle=270]{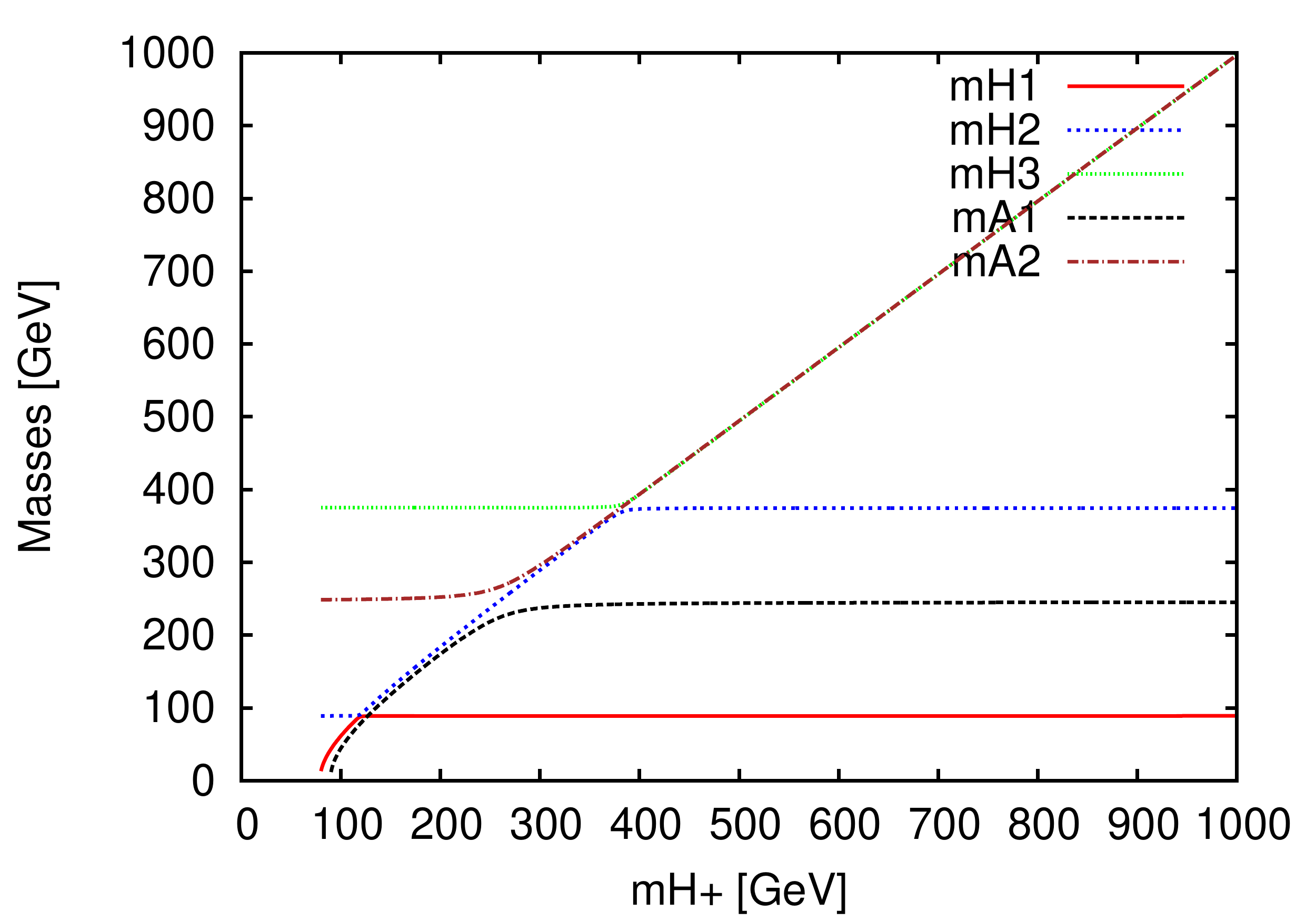}}
	\subfloat[Large couplings]{\includegraphics[width= 60 mm, angle=270]{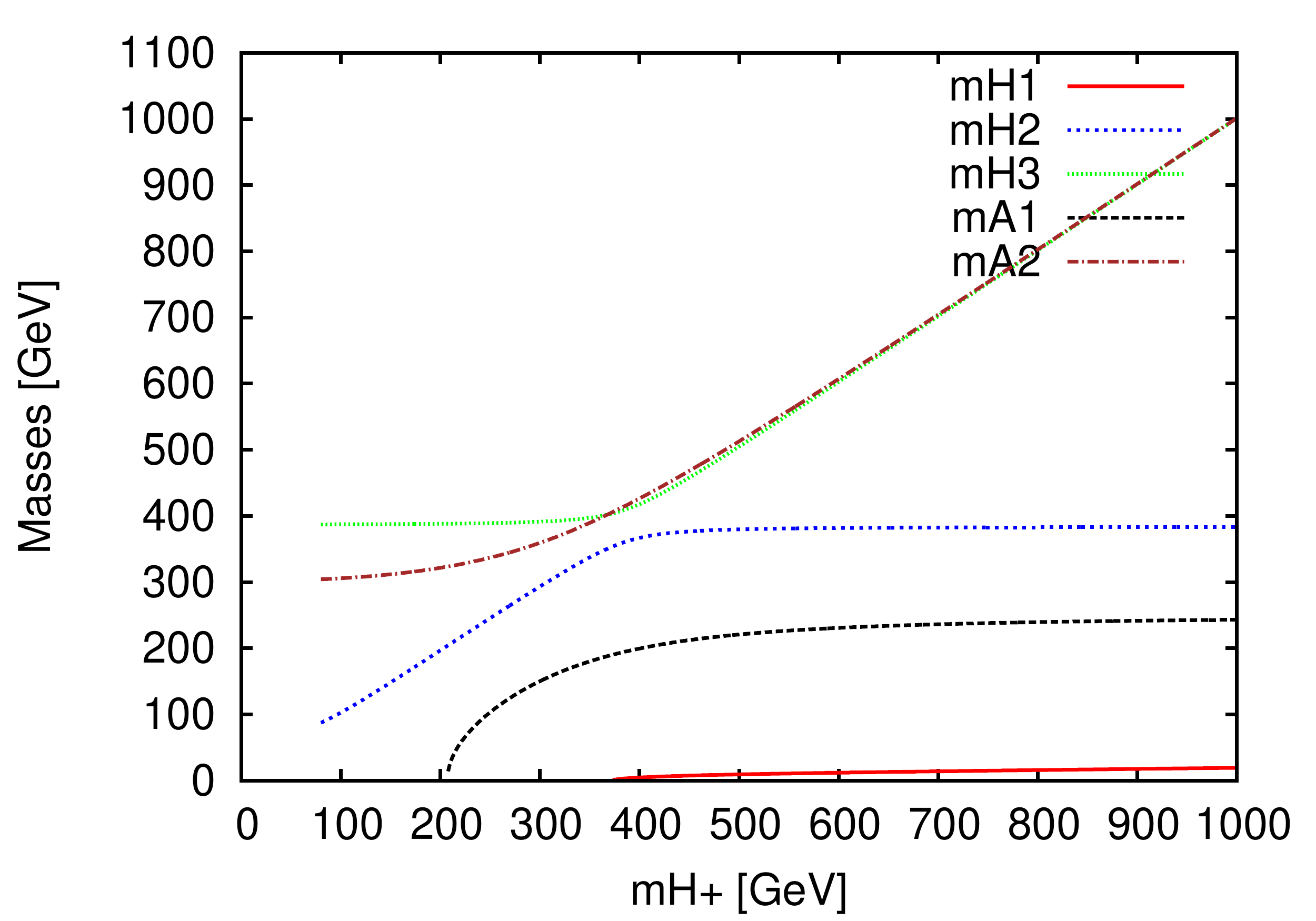}}

\caption{The Higgs masses as a function of $m_{H^\pm}$ for $\tan \beta = 35$. In (a),we have $\kappa = 0.1, \lambda = 0.1$, while in (b) $\kappa = 0.5, \lambda = 0.5$. In both cases $A_\kappa = -100$ GeV.}
\label{mChargedtanbeta30}
\end{figure}

What happens if you increase the value of $\tan \beta$, making one of the doublet VEVs much larger than the other? In figure \ref{mChargedtanbeta30}, this is shown as a function of the charged Higgs mass for two different choices of the singlet couplings. From the figure we see that there seems to be rather distinct points where the identity of two different Higgses seem to switch. For example, the $H_1$ and the $H_2$ states seem to switch behaviour with respect to the charged Higgs mass at around $m_{H^\pm} = 120$ GeV. This kind of behaviour really comes from the way the mixing matrices depends on $m_{H^\pm}$ (or equivalently $A_\lambda$) and from how we label the different  states. A large $\tan \beta$ value means that the mixing with the $H_u$ doublet will be much more important in terms of mass than the mixing with the $H_d$ doublet. From this we can understand why a larger $\tan \beta$ value makes the identities of the Higgses more sharply defined.

What is perhaps more interesting to note is that we have three masses here that are almost independent of the charged Higgs mass, even though exactly what we call the state varies with $m_{H^\pm}$. This is also coupled to the fact that the couplings in this plot is rather small, $\kappa = \lambda = 0.1$. If we instead make them larger, we instead get the behaviour seen in figure \ref{mChargedtanbeta30}, where the ``switching'' behaviour is not at all as sharp. Increasing the singlet couplings also pushes the lowest physically allowed value for $m_{H^\pm}$ upwards (i.e.\ the value where $m_{H_1} > 0$), and seems to give the lightest scalar Higgs a very small mass. 

\subsection{The MSSM limit} \label{sec:MSSMLimit}
A few things can be noted analytically when we take the limit $\lambda, \kappa \rightarrow 0$ while keeping $\kappa / \lambda = k$ fixed. For example, we can see that the parameter $m_S^2$ in the Lagrangian will approach a fixed value as soon as $\lambda$ and $\kappa$ get small. This is seen by looking at the expression for $m_S^2$ we got from the requirement of vacuum stability, equation (\ref{m_sExpression}). Since we keep $\mu = \frac{1}{\sqrt{2}}v_s \lambda$ fixed, we have that 
\begin{eqnarray*}
 m_S^2 &=& - \frac 1 2 v^2 \lambda^2 - v_s^2 \kappa^2 + \frac{1}{\sqrt{2}} A_\lambda \lambda \frac{v_u v_d}{v_s} + v_u v_d \kappa \lambda - \frac{1}{\sqrt{2}}A_\kappa v_s \kappa  \\
&=&   - \frac 1 2 v^2 \lambda^2 - 2\mu^2 k^2 + \frac{1}{2} A_\lambda \lambda^2 \frac{v_u v_d}{\mu} + v_u v_d k \lambda^2 -  A_\kappa k \mu \\ 
&\rightarrow &  k \mu(A_\kappa-k\mu)  \mbox{ as } \lambda \rightarrow 0, 
\end{eqnarray*}
also using the relation between $\kappa$ and $\lambda$. If we instead keep $\kappa$ fixed and let $\lambda \rightarrow 0$, we see that instead
\begin{eqnarray*}
m_S^2 &=&  - \frac 1 2 v^2 \lambda^2 - v_s^2 \kappa^2 + \frac{1}{2} A_\lambda \lambda^2 \frac{v_u v_d}{\mu} + v_u v_d \kappa \lambda - A_\kappa \kappa \frac{\mu}{\lambda} \\
&\rightarrow& \infty \mbox{,  as  } \lambda \rightarrow 0.
\end{eqnarray*} 

\begin{figure}

	\centering
	\includegraphics[width= 70 mm, angle=270]{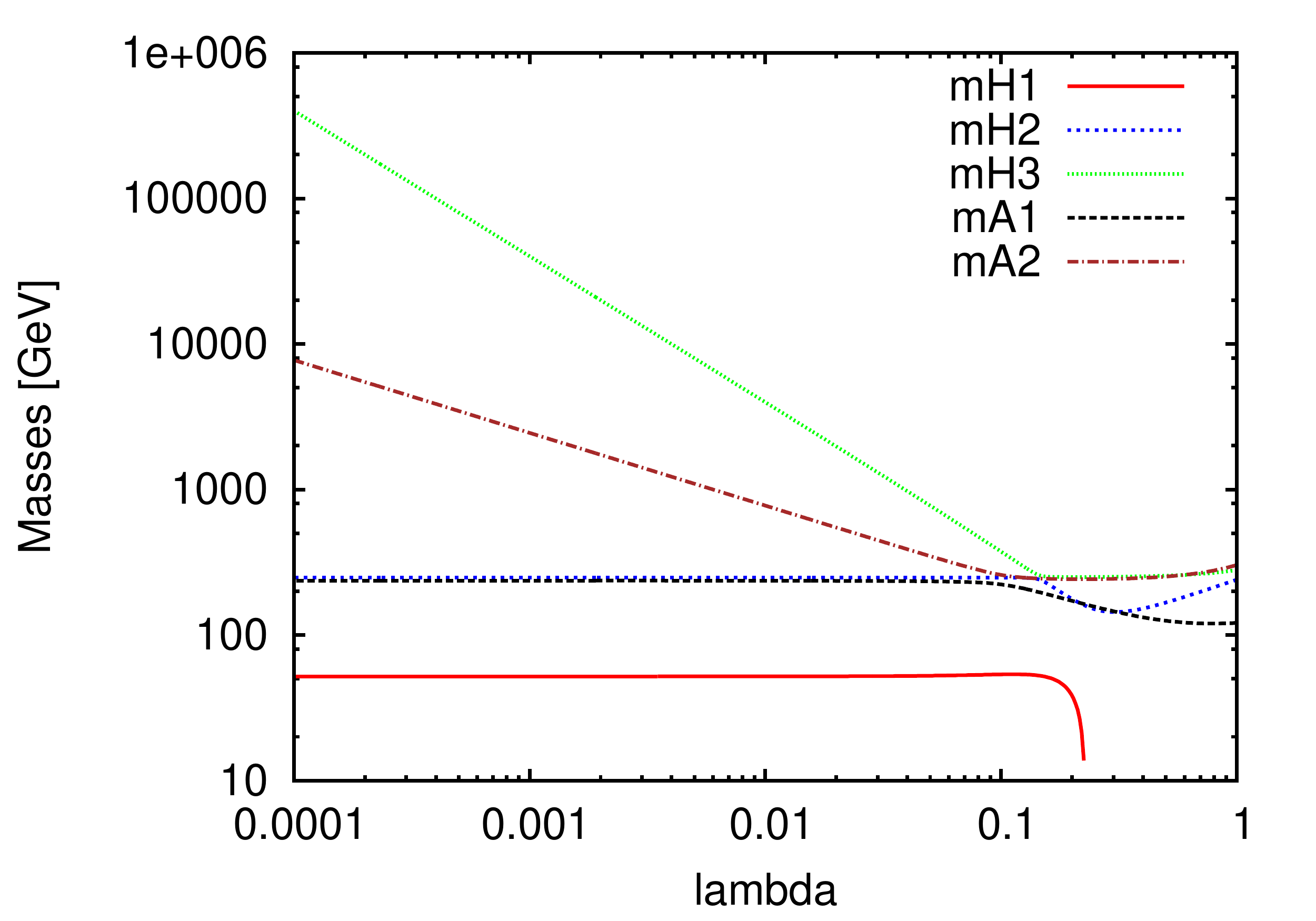}

\caption{How the masses vary when $\lambda$ goes to zero while keeping $\kappa$ fixed. In this plot, $\tan \beta = 2, m_{H^\pm} = 250 \mbox{ GeV}, \kappa = 0.1$ and $A_\kappa = -100 $ GeV. }
\label{lambdalimit}
\end{figure}

This behaviour is shown in figure \ref{lambdalimit}, and it means that just as we can see from looking at the Lagrangian (\ref{NMSSM-Lagrangian}), as $\lambda$ goes to zero the singlet field decouples and the masses of the singlet dominated states blow up.

\begin{figure}

	\centering
	\includegraphics[width= 70 mm, angle=270]{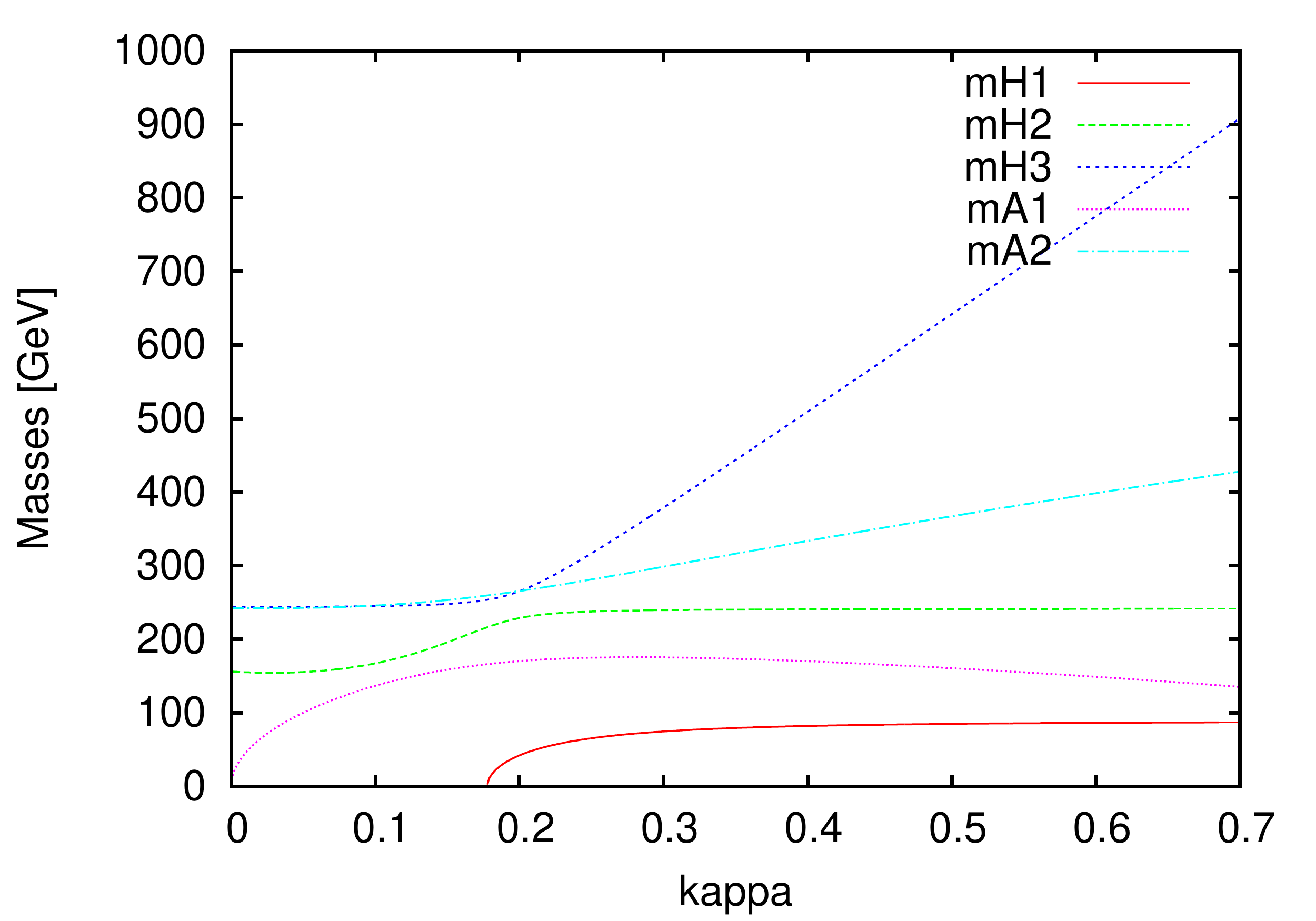}

\caption{The masses as a function of $\kappa$, where $\tan \beta = 7, m_{H^\pm} = 250 \mbox{ GeV}, A_\kappa = -100$ GeV and $\lambda = 0.3 $. }
\label{runningKappa}
\end{figure}

Another way of taking an interesting limit is to let $\kappa \rightarrow 0$ keeping $\lambda$ constant, as can be seen in figure \ref{runningKappa}. Here we see that in this limit (which really isn't a proper MSSM limit since $\lambda$ stays large and thus the singlet doesn't fully decouple), the $A_1$ state becomes massless, which again is because we restore the PQ symmetry turning $A_1$ into the massless PQ-axion. However, it is seen that for all cases with fixed $\lambda$ there is no way of keeping $H_1$ at a positive mass squared as $\kappa$ goes to zero. For smaller values of $\lambda$, $m^2_{H_1}$ becomes negative for smaller values of $\kappa$, but any given value of $\lambda$ ultimately restricts the lowest possible value of $\kappa$. So from looking at this in addition to the above discussed $\lambda \rightarrow 0$ limit we see that in order to decouple the singlet and reduce the theory to MSSM, one is in effect forced to take the simultaneous limit $\kappa,\lambda \rightarrow 0$ (or in addition take the limit $A_\kappa\rightarrow 0$ as studied later).

\begin{figure}

	\centering
	\subfloat[Masses]{\includegraphics[width= 6 cm, angle=270] {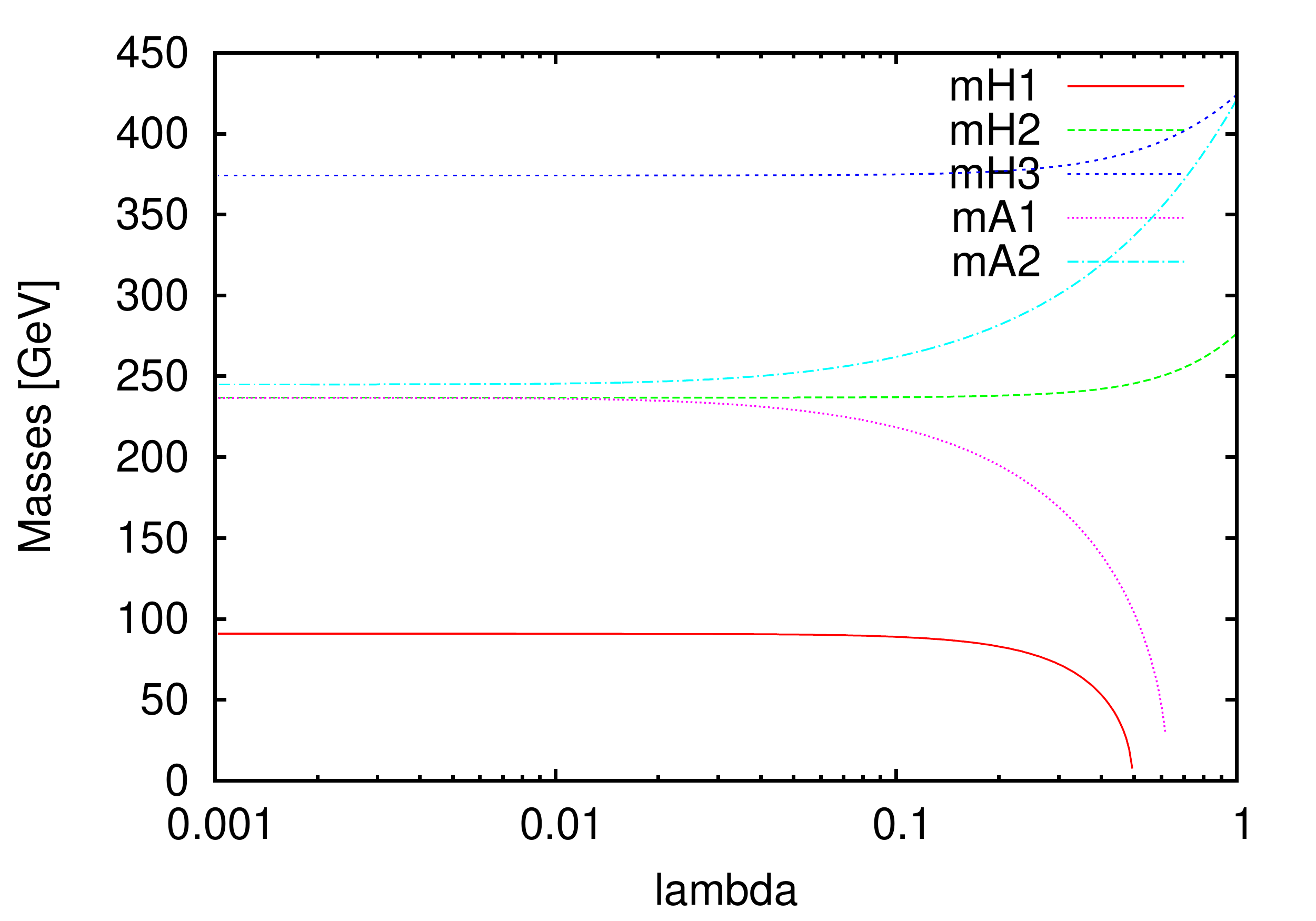}}
	\subfloat[$H_iVV$-couplings and $\cos\theta_A$] {\includegraphics[width= 6 cm, angle=270] {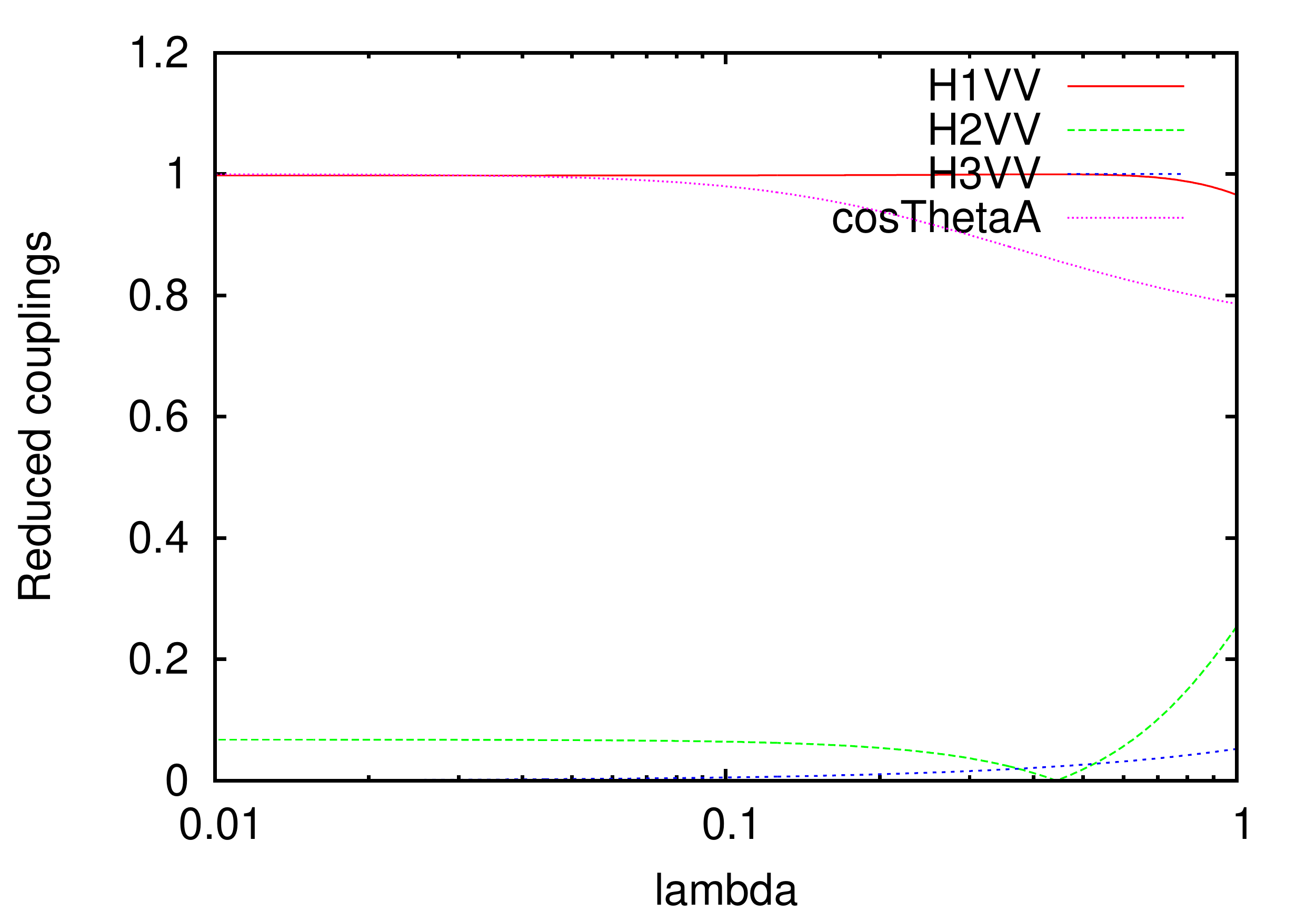}}

\caption{ The masses, $H_iVV$-couplings and $\cos \theta_A$ as the MSSM limit is approached, using $k=1$, $\tan \beta = 2, m_{H^\pm} = 250 $ GeV and $A_\kappa = -100$.}

\label{mssmlimit1}
\end{figure}

In figure \ref{mssmlimit1}, we see that when approaching the MSSM limit in the sensible way, none of the masses becomes large, which is understandable since we above showed that $m_S^2$ tends towards a finite value. However, also in this case the singlet decouples and does not mix with the other fields, which we see by looking at the reduced couplings and $\cos \theta_A$, seeing that the lightest Higgs state becomes completely standard model like, and also $\cos \theta_A$ goes to 1, so that both mixing matrices become block diagonal and there is no mixing between the doublet and the singlet. That the lightest scalar becomes standard model like in this limit means that for small $\kappa$ and $\lambda$ of roughly the same size ($k \sim 1$) means that it could be detected as easily as in the standard model.

We also see that $\cos \theta_A$ goes to 1 much slower than the reduced scalar couplings. This is not a general feature but depends on the value of $k$. However, I've not found any cases with $k>1$ where the pseudoscalar mixing disappears slower than the scalar mixing. This means that small values of $\lambda$ and $\kappa$ suppresses mixing between scalar singlet and doublet states much more than between the pseudoscalar states. Moreover, as $\lambda$ becomes small the mixing between the scalar Higgs fields stop depending on $\lambda$ and stays more or less constant. This is of course very reasonable since $\lambda$ determines the coupling strength between the singlet and the doublets.

\begin{figure}

\centering
	\subfloat[Masses]{\includegraphics[width= 6 cm, angle=270] {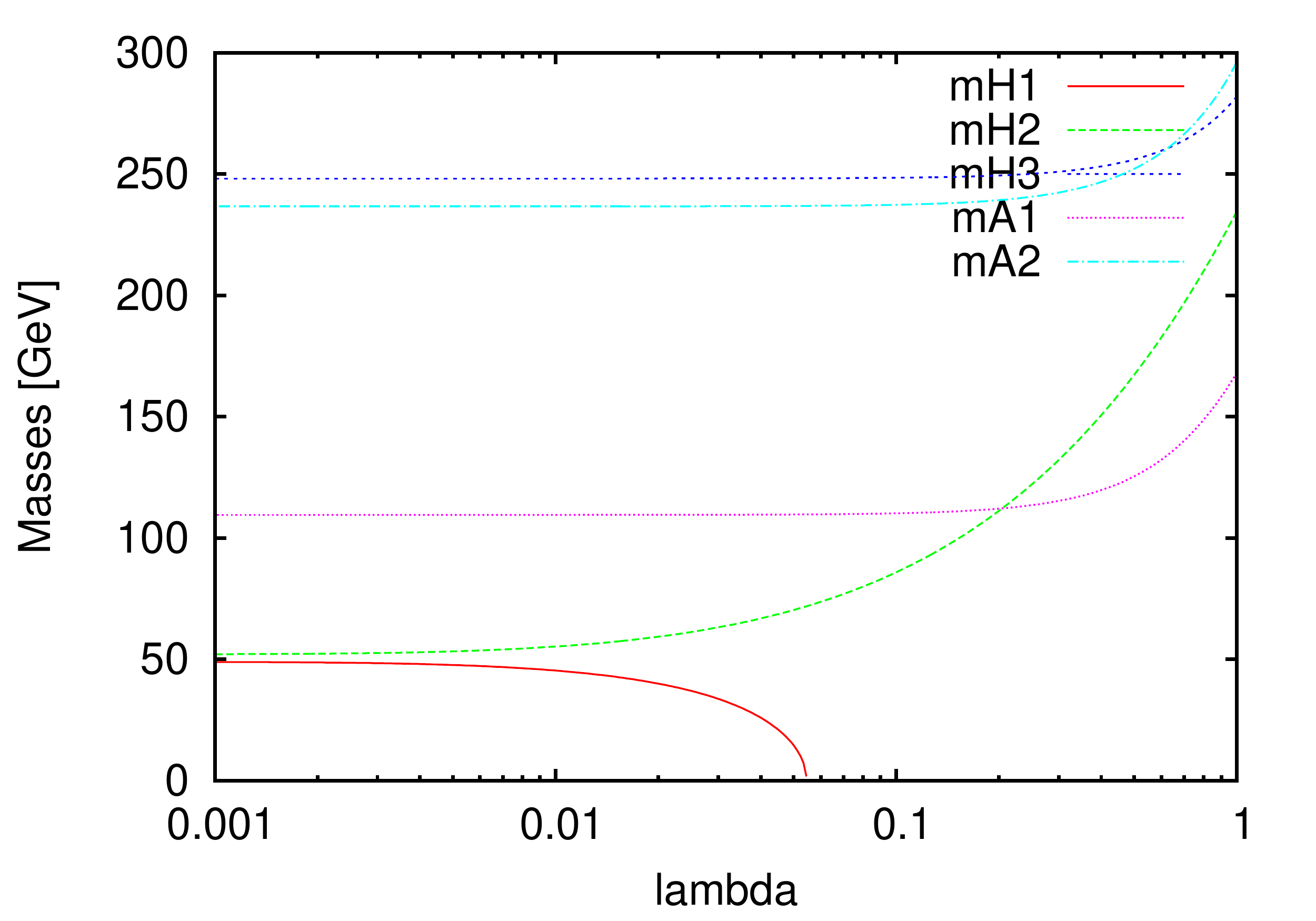}}
	\subfloat[Couplings and $\cos\theta_A$] {\includegraphics[width= 6 cm, angle=270] {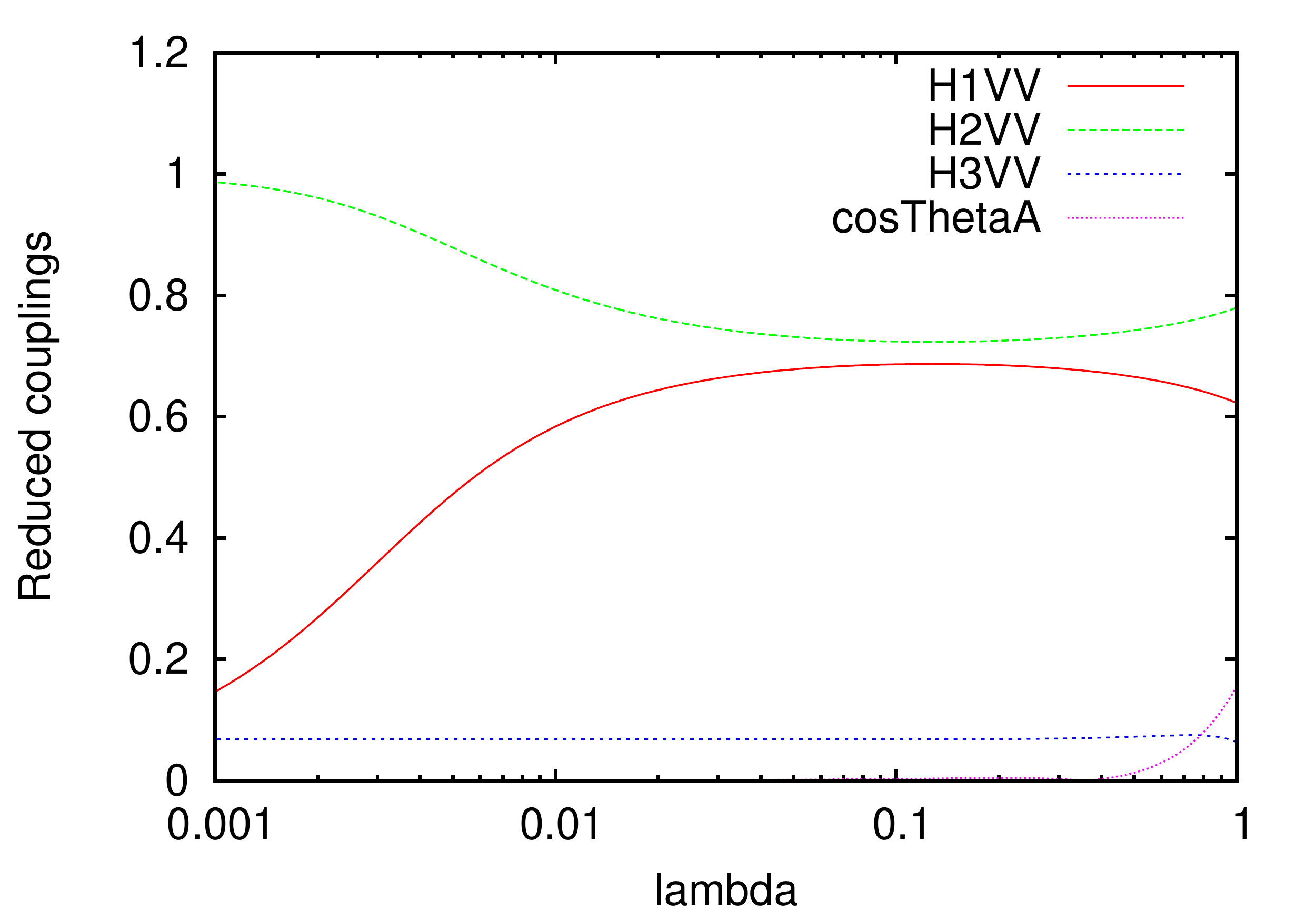}}

\caption{How the masses, $\cos \theta_A$ and $H_iVV$ couplings vary when approaching the MSSM limit, using $k=0.2, \tan \beta = 2, m_{H^\pm} = 250$ GeV and $A_\kappa = -100$ GeV.}
\label{mssmlimit-rapid}
\end{figure}
However, as we can see in figure \ref{mssmlimit-rapid}, the reduced couplings (and thus the mixing) doesn't always vary slowly when we take the limit. This behaviour seems to occur only when $k \lesssim 0.5$, in the figure we have $k=0.2$ as a representative case. In these scenarios, the couplings continue to vary very rapidly (considering the logarithmic scale) even when $\lambda$ is very small, and the mixing only disappears when $\lambda$ becomes really small. Differently from before, the lightest Higgs isn't the most standard-model like in this scenario. This role is instead filled by $H_2$. We also note that the masses of $H_1$ and $H_2$ gets very similar in the MSSM limit.

Since $\cos \theta_A$ approaches 0, we see that the lightest pseudoscalar state also decouples in the MSSM limit (that $\cos \theta_A = 0$ of course means that the off-diagonal element of the pseudoscalar mixing matrix $\sin \theta_A=1$).

We however see that the masses of the two lightest Higgses are very small for these parameter choices, so this particular case is not realistic. For such small values of $k$ the requirement that $m^2_{H_1} >0$ seems to rule out most of the parameter space, i.e.\ for many choices of other parameters, $m^2_{H_1}<0$. I've not found any case where the same thing happens for the pseudoscalar mixing, but no methodical search of such a scenario was carried out.

\begin{figure}

\centering
	\subfloat[$k=2$]{\includegraphics[width= 6 cm, angle=270] {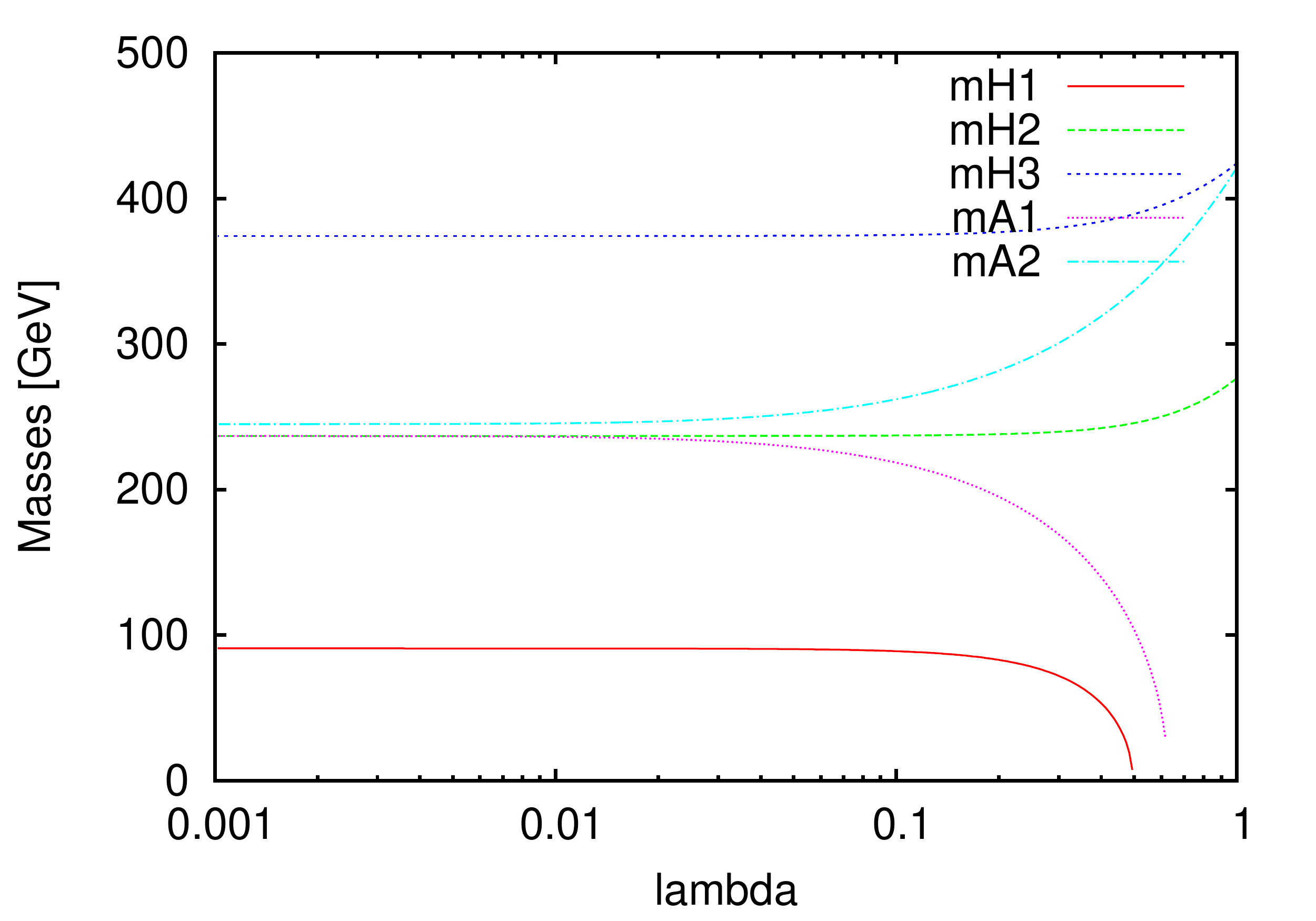}}
	\subfloat[$k=10$] {\includegraphics[width= 6 cm, angle=270] {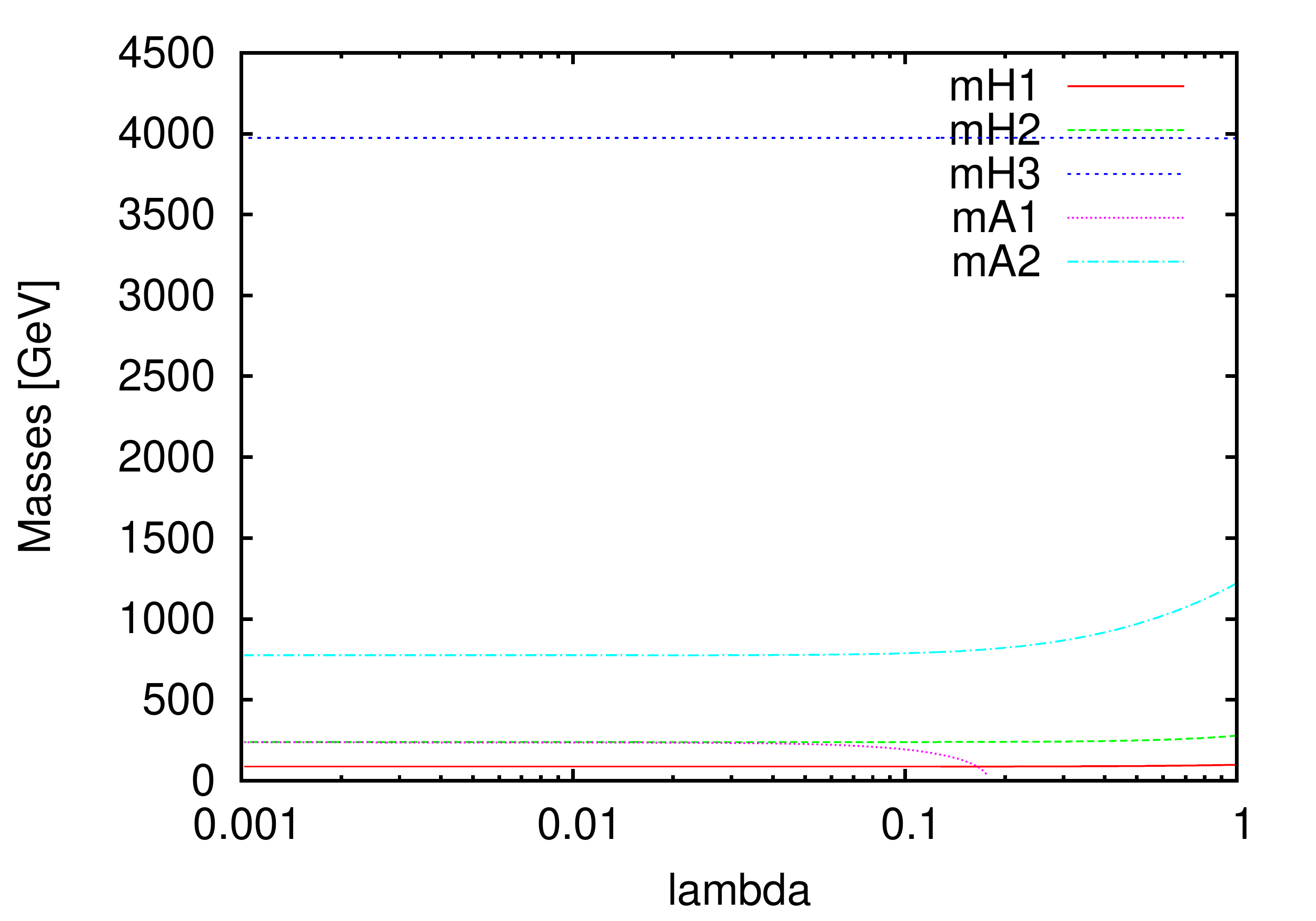}}

\caption{How the masses vary when approaching the MSSM limit for two different values of $k$. The rest of the parameters have the same value as those in figure \ref{mssmlimit1}.}
\label{mssmlimit2}
\end{figure}

\begin{figure}

	\centering
	\subfloat[Masses]{\includegraphics[width= 60 mm, angle=270]{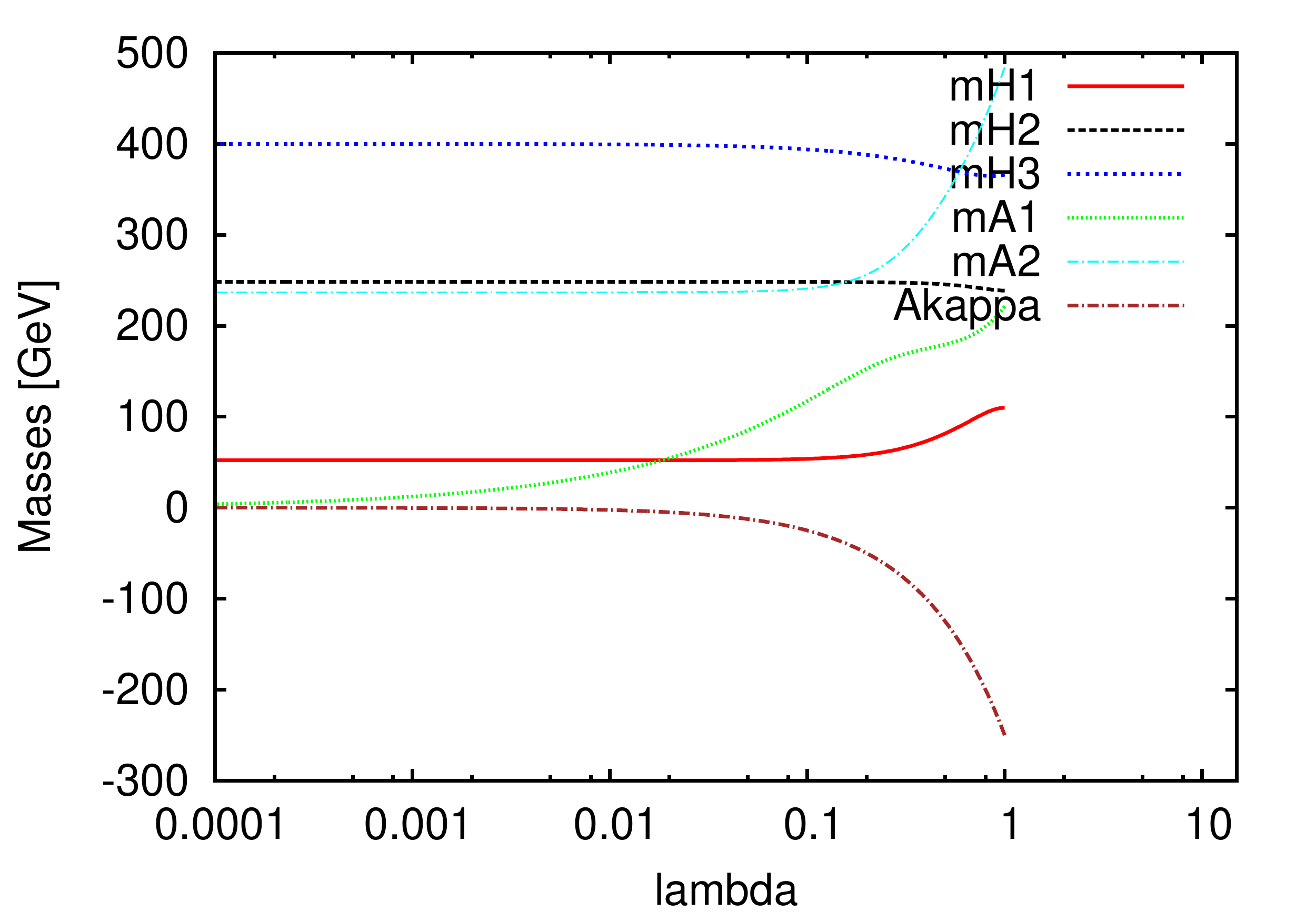}}
\subfloat[$H_iVV$ couplings and $\cos \theta_A$]{\includegraphics[width= 60 mm, angle=270]{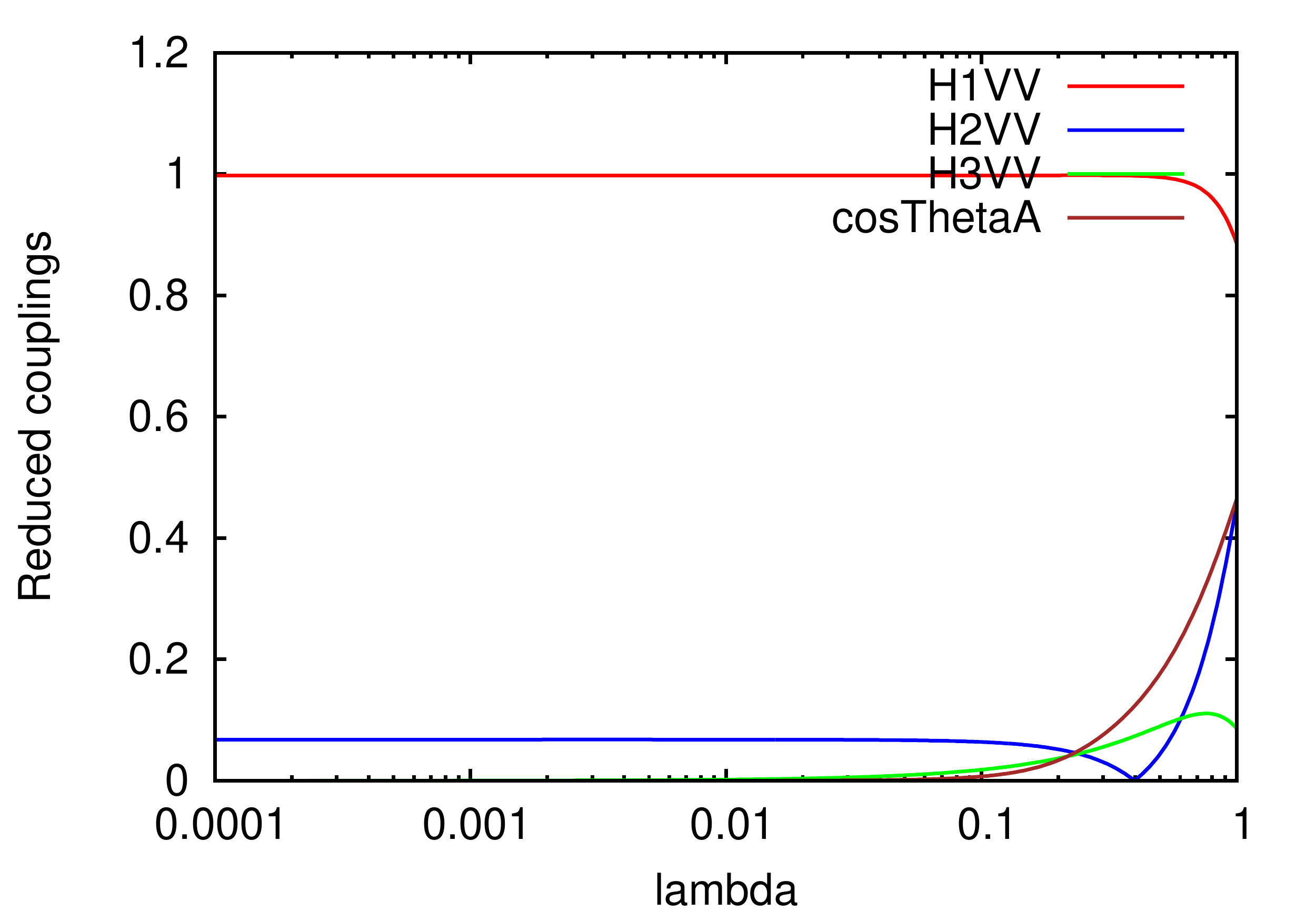}}

\caption{How the masses, couplings and $\cos \theta_A$ vary when $\lambda$ goes to zero while keeping $k = \kappa/\lambda$ fixed, in addition to also sending $A_\kappa$ to zero. In this plot, $k = 1, \tan \beta = 2, m_{H^\pm} = 250 \mbox{ GeV}$. $A_\kappa$ begins on a value of $-250$ GeV, and is sent to zero in the same manner as $\lambda$ and $\kappa$ (i.e.\ we keep the ratio $\kappa / A_\kappa$ constant).}
\label{mssmlimitAkappa}
\end{figure}

As we can see from the formula for $m_S^2$ in the MSSM limit, if we let the ratio $k=\kappa /\lambda$ get larger, then the masses of the singlet states should increase. And this is exactly what happens as we can see in figure \ref{mssmlimit2}, where a larger $k$ value is seen to push the heaviest scalar, the singlet dominated one, far up in mass, whereas the mass of the heavy pseudoscalar is also increased but not at all as much.

Another possible way of approaching the MSSM limit is to also send the $A_\kappa$ parameter to zero, which effectively further suppress the $S^3$ term in the Lagrangian and thus restores the PQ-symmetry. And as we can see in figure \ref{mssmlimitAkappa}, in this case the lightest pseudoscalar Higgs become massless in the limit, restoring the massless PQ axion. In this limit, we also see that the pseudoscalar mixing disappears ($\cos \theta_A \rightarrow 1$) and that the lightest scalar Higgs again is fully standard model like.

One could also imagine letting $v_s$ go to zero smoothly, but in order to keep $\mu$ in an acceptable range this would mean that $\lambda$ would have to become larger than allowed by the requirement of perturbation theory being valid up to the GUT-scale, i.e.\ $\lambda \lesssim 0.7$ as stated before. A small value of $\mu$ is excluded from bounds on the chargino masses\cite{ParticleDataGroup} from direct searches. As said towards the end of section \ref{NMSSM-section}, we can give up the perturbative requirement and consider $\lambda \gg 1$ as in the $\lambda$-SUSY model \cite{Barbieri:2006bg}, in which case this limit could be viable.

\section{Summary and conclusions}
In this study, we have briefly reviewed some of the motivation behind studying supersymmetry. Then the MSSM was described, and its simplest extension, the NMSSM was introduced as a solution to the $\mu$-problem. Some technical details of the Higgs sector of NMSSM was stated, including the mass matrices and definition of the reduced couplings. This was followed by some numerical studies of different choices of parameters, including a look at approaching the MSSM limit in some different ways.

We find that in the by renormalization group flow favoured choice of $\kappa,\, \tan \beta$ and $\lambda$, the mass spectra with three different relatively light Higgs bosons should make the theory easy to distinguish from the MSSM even when not all the Higgses are detected. But in other perfectly allowed cases, the distinction might not be directly obvious. We also see that for this case some of the couplings to SM-particles pass through 0, so that it is possible that for example the $H_2$ state can be hidden and not interact in a standard model like way at all.

For a larger $\kappa$ value, i.e.\ a more strongly broken PQ-symmetry, we see that the couplings behave in a qualitatively different way. In this case the switching behaviour takes place inside the physically sensible area, but the couplings of the $H_1$ state doesn't show any complicated dependence on $m_{H^\pm}$. Over the whole range, $H_1$ is the essentially fully standard model like.

If we want to study the limits where the singlet decouple, from looking at what happens when only one of $\kappa$ or $\lambda$ are sent to zero, we conclude that sensible limits exists only when both of them are decreased simultaneously. When we approach this MSSM limit, keeping the ratio $\kappa/\lambda$ constant, the mixing with the singlet field disappears. Mostly this decoupling happens rather quickly, but in some cases, when $\kappa/\lambda \lesssim 0.5$, the mixing of the scalar states depends strongly on the nonzero $\lambda$. We also see that in this case, the lightest Higgs is no longer necessarily the most standard model like, that role being taken by $H_2$. But for $\lambda > 0$, both $H_1$ and $H_3$ are slightly standard model like, so if this is the case (i.e.\ we have a small $\lambda$ and a ratio as described), we could detect three relatively light Higgs bosons with different masses. However, these cases seem to depend very much on the ratio $\kappa/\lambda$ having a specific value, and also seem to give the lightest states too low masses for it to be realistic, but since I have not scanned all of parameter space and in addition I am only doing tree level calculations, this kind of scenario cannot be altogether ruled out.

So we see from all this that the NMSSM model offers many interesting possibilities not seen in the MSSM.

\end{document}